\documentclass[12pt]{article}
\usepackage{amsmath,amssymb,latexsym,epsfig,psfrag}
\allowdisplaybreaks

\def\nslash{\rlap{\hspace{0.02cm}/}{n}}
\def\nbslash{\rlap{\hspace{0.02cm}/}{\bar n}}
\def\ff{f\hspace{-0.3cm}f}
\def\muf{\mu_f}
\def\muh{\mu_h}
\def\mui{\mu_s}

\begin{document}

\begin{titlepage}

\begin{flushright}
CLNS~07/2009\\
FERMILAB-PUB-07-507-T\\
MZ-TH/07-16\\[0.2cm]
October 2, 2007
\end{flushright}

\vspace{0.2cm}
\begin{center}
\Large\bf
Dynamical Threshold Enhancement and Resummation in Drell-Yan Production
\end{center}

\vspace{0.2cm}
\begin{center}
{\sc Thomas Becher$^a$, Matthias Neubert$^{a,b,c}$ and Gang Xu$^{c}$}\\
\vspace{0.4cm}
{\sl $^a$\,Fermi National Accelerator Laboratory\\
P.O. Box 500, Batavia, IL 60510, U.S.A.\\[0.3cm]
$^b$\,Institut f\"ur Physik (THEP), Johannes Gutenberg-Universit\"at\\ 
D--55099 Mainz, Germany\\[0.3cm]
$^c$\,Institute for High-Energy Phenomenology\\
Newman Laboratory for Elementary-Particle Physics, Cornell University\\
Ithaca, NY 14853, U.S.A.}
\end{center}

\vspace{0.2cm}
\begin{abstract}
\vspace{0.2cm}
\noindent 
Partonic cross sections for the production of massive objects in hadronic collisions receive large corrections when the invariant mass of the initial-state partons is just above the production threshold. Since typically the center-of-mass energy of the hadronic collision is much higher than the mass of the heavy objects, it is not obvious that these contributions translate into large corrections to the hadronic cross section. Using a recent approach to threshold resummation based on effective field theory, we quantify to which extent the fall-off of the parton densities at high $x$ leads to a dynamical enhancement of the partonic threshold region. With the example of Drell-Yan production, we study the emergence of an effective physical scale characterizing the soft emissions in the process. We derive compact analytical expressions for the resummed Drell-Yan cross section and rapidity distribution directly in momentum space. They are free of Landau-pole singularities and are trivially matched onto fixed-order perturbative calculations. Evaluating the resummed cross sections at NNNLL order and matching onto NNLO fixed-order calculations, we perform a detailed numerical analysis of the cross section and rapidity distribution in $pp$ collisions.
\end{abstract}
\vfil

\end{titlepage}

\section{Introduction}

The Drell-Yan process \cite{Drell:1970wh}, the production of a lepton pair in hadron-hadron collisions, has played an important role in establishing the parton picture underlying the description of hard interactions in QCD. In current experiments, studies of the Drell-Yan cross section as a function of the invariant mass of the lepton pair are used to search for new heavy particles such as a hypothetical $Z'$ boson, while the differential distributions provide detailed information about the parton distribution functions (PDFs), including in particular the sea-quark distributions. Drell-Yan production also serves as a prototype for other collider processes, such as Higgs production or the production of new particles. 

A lot of effort has been put in obtaining accurate theoretical predictions for the Drell-Yan process in perturbative QCD. The calculation of the cross section and rapidity distribution at next-to-leading order (NLO) in $\alpha_s$ was  accomplished in the pioneering work \cite{Altarelli:1979ub}. The first next-to-next-to-leading order (NNLO) result for the cross section was presented in \cite{Hamberg:1990np} and confirmed much later in \cite{Harlander:2002wh}. The rapidity distribution at NNLO was derived in \cite{Anastasiou:2003yy,Anastasiou:2003ds}, while the fully differential cross section was obtained recently in \cite{Melnikov:2006di,Melnikov:2006kv}.

As the invariant mass $M$ of the lepton pair approaches the center-of-mass energy of the collision, there is less and less phase space available for the emission of QCD radiation. After the cancellation of virtual and real soft divergences large Sudakov logarithms remain, because the scale associated with the soft radiation is much smaller than $M$. These ``threshold logarithms" threaten the convergence of the perturbative expansion and need to be resummed to all orders. For the inclusive cross section $d\sigma/dM^2$ this was accomplished in the seminal papers \cite{Sterman:1986aj,Catani:1989ne} based on the solution of certain evolution equations in Mellin moment space (see also \cite{Magnea:1990qg,Korchemsky:1993uz}). The generalization of this method to the Drell-Yan rapidity distribution was obtained in \cite{Sterman:2000pt}. Recent analyses of threshold resummation for the rapidity distribution in Drell-Yan or electroweak gauge boson production can be found in \cite{Mukherjee:2006uu,Bolzoni:2006ky} in next-to-leading logarithmic (NLL) approximation, while in \cite{Ravindran:2006bu,Ravindran:2007sv} the resummation is extended to the next-to-next-to-next-to-leading logarithmic (N$^3$LL) order. 

Because the PDFs are strongly suppressed in the endpoint region $x\to 1$, the cross section $d\sigma/dM^2$ is a  steeply falling function as $M$ approaches the kinematical endpoint $\sqrt{s}$. In fact, in a typical experiment it will not be possible to observe Drell-Yan pairs with masses exceeding about one half of the center-of-mass energy. For instance, at the LHC one does not expect to discover new heavy particles with masses in the 10\,TeV range. In practice, one is therefore never in a region where the ratio $\tau=M^2/s$ approaches 1. Since threshold resummation deals with logarithms of the form $\ln(1-\tau)$, it is then not obvious why such terms should be treated on different footing than other higher-order terms. In view of this, it is surprising that large resummation effects were recently claimed to be important for the Drell-Yan rapidity distribution as measured by the E866/NuSea collaboration \cite{Webb:2003ps}. Specifically, Ref.~\cite{Bolzoni:2006ky} claims that for $\sqrt{s}=38.76$\,GeV and $M=8$\,GeV (corresponding to $\tau\approx 0.04$) the resummation of threshold logarithms at NNLL order would lower the cross section by about 30\% compared with the fixed-order NLO result, whereas the fixed-order NNLO corrections increase it by a small amount \cite{Anastasiou:2003yy,Anastasiou:2003ds}. However, significantly smaller resummation effects were found by other authors for LHC kinematics ($M=115$\,GeV and $\sqrt{s}=14$\,TeV) \cite{Ravindran:2006bu} and for $W$-boson production at RHIC ($M=80.4$\,GeV and $\sqrt{s}=500$\,GeV) \cite{Mukherjee:2006uu}. An argument why threshold resummation effects could be important even if $\tau\ll 1$ has been given in Refs.~\cite{Appell:1988ie,Catani:1998tm}. The idea is that the sharp fall-off of the parton luminosity at large $x$ dynamically enhances the contribution of the {\em partonic\/} threshold region $z=M^2/\hat s\to 1$, i.e., the region where the center-of-mass energy $\sqrt{\hat s}$ of the initial-state partons is just sufficiently large to produce the Drell-Yan pair. It could then be important to resum logarithms of the form $\ln(1-z)$ in the hard partonic cross section. However, since $(1-z)$ is not related to a small ratio of external physical scales, it is not obvious how to give a formal justification of this argument. 

To study this question quantitatively and to assess the importance of resummation effects for the Drell-Yan rapidity distribution were the main motivations for the present work. To do so, we use a recent approach to Sudakov resummation based on effective field theory \cite{Becher:2006nr,Becher:2006mr}. Contrary to the standard treatment in Mellin moment space, this framework completely separates the effects associated with different scales in the problem, thereby avoiding the Landau-pole ambiguities inherent in the standard approach. It then uses renormalization-group (RG) evolution to resum logarithms of scale ratios. The resummation is performed directly in momentum space, which makes it simpler to compare to and match onto fixed-order calculations. Our framework is particularly well suited to study the resummed rapidity distribution, for which we derive an exact analytic expression as a one-dimensional integral over PDFs. Using the convergence properties of the perturbative expansion after scale separation as the primary criterion, we study in detail how and under which circumstances an effective physical scale $\mui\ll M$ emerges, which is associated with the soft emission in the process. Our approach resums logarithms of the ratio $M/\mui$ to all orders in perturbation theory. 

Our main findings can be summarized as follows: 
\begin{itemize}
\item[i.]
In the true endpoint region $\tau\to 1$, the effective soft scale $\mui$ is an order of magnitude smaller than the naive guess $M(1-\tau)$. For PDFs behaving like $f_{i/N}(x)\sim(1-x)^{b_i}$ near $x\to 1$, we find $\mui\approx\lambda^{-1} M(1-\tau)$ with $\lambda\approx 2+b_q+b_{\bar q}=O(10)$. This result provides a formal justification to the argument of a dynamical enhancement of the partonic threshold region due to the fall-off of parton densities. 
\item[ii.]
The dynamical enhancement of the threshold contributions remains effective down to moderate values $\tau\approx 0.2$, while at very small $\tau$ values the parameter $\lambda$ decreases to about 2. This reflects the fact that for small $x$ values the fall-off of the PDFs is much weaker than for large $x$.
\item[iii.]
Even far away from the true threshold the Drell-Yan cross section receives its dominant contributions from those terms in the hard partonic cross section that are leading in the limit $z\to 1$. Assuming this is true for other processes as well, the evaluation of virtual corrections plus soft emissions provides a simple and efficient way to obtain useful approximations for higher-order perturbative corrections.
\item[iv.]
With the appropriate choice of the effective soft scale $\mui$, the convergence of the perturbative expansion is greatly improved by the resummation. However, for small Drell-Yan masses the terms beyond $O(\alpha_s^2)$ in the resummed expression for the cross section are numerically unimportant. We thus do not confirm the large impact of threshold resummation on the Drell-Yan rapidity distribution reported in \cite{Bolzoni:2006ky}. For larger masses the effects can be significant. For instance, the experiment E866/NuSea has reported data up to $M=16.85$\,GeV (corresponding to $\tau\approx 0.19$) \cite{Webb:2003ps}. We find that at $M=16$\,GeV resummation effects enhance the fixed-order predictions for the cross section by about 25\% at NLO, and 7\% at NNLO.
\item[v.]
For the case of the integrated cross-section $d\sigma/dM^2$, we perform a detailed comparison with the traditional resummation approach in moment space. Similar to the case of deep-inelastic scattering (DIS) in the region $x\to 1$, we find that the two approaches are equivalent up to power corrections, which turn out to be numerically small. An important conceptual difference is that in the effective-theory approach the running coupling is evaluated at physical short-distance scales depending only on the external variables $s$ and $M$ (and perhaps the rapidity $Y$). In this way, the Landau-pole ambiguities inherent in the standard approach are avoided. 
\end{itemize}

We begin our analysis discussing the structure of the hard-scattering kernels relevant for the Drell-Yan rapidity distribution in fixed-order perturbative QCD. In Section~\ref{sec:SCET} we use the framework of soft-collinear effective theory (SCET) \cite{Bauer:2000yr,Bauer:2001yt,Beneke:2002ph} to derive the standard factorization formula for the partonic cross section in the limit $z\to 1$ in terms of hard and soft functions, which we define in terms of Wilson coefficients of operators in the effective theory. The solutions to the RG equations obeyed by these coefficients are derived in Section~\ref{sec:resummation}. With these results at hand, we present exact analytic expressions for the resummed Drell-Yan cross section and rapidity distribution. A detailed numerical analysis of our results is presented in Section~\ref{sec:numerics}. After choosing the hard and soft matching scales in the effective theory by analyzing the perturbative expansions of the Wilson coefficient functions, we investigate the stability of the results under scale variations and discuss the impact of the resummation. Before concluding, we discuss the connection with the conventional moment-space approach.

\section{Fixed-order calculation and the threshold region}
\label{sec:pQCD}

We consider the production of a lepton pair with invariant mass $M$ in hadron-hadron collisions at center-of-mass energy $\sqrt s$ (Drell-Yan process), focusing for simplicity on the reaction $N_1+N_2\to\gamma^*+X$ followed by $\gamma^*(q)\to l^-+l^+$. Our goal is to calculate the double differential cross section in the variables $M^2=q^2$ and $Y=\frac12\ln\frac{q^0+q^3}{q^0-q^3}$, where $Y$ denotes the rapidity of the lepton pair in the center-of-mass frame. Up to power corrections this cross section can be calculated in perturbative QCD and expressed in terms of convolutions of short-distance partonic cross sections with PDFs:
\begin{equation}\label{sig}
   \frac{d^2\sigma}{dM^2 dY}
   = \frac{4\pi\alpha^2}{3N_c M^2 s} \sum_{i,j} \int dx_1\,dx_2\,
   \widetilde C_{ij}(x_1,x_2,s,M,\muf)\,
   f_{i/N_1}(x_1,\muf)\,f_{j/N_2}(x_2,\muf) \,.
\end{equation}
Here $f_{i/N}(x,\muf)$ is the probability of finding a parton $i$ with longitudinal momentum fraction $x$ inside the hadron $N$, and $\muf$ is the factorization scale. The hard-scattering kernels $\widetilde C_{ij}$ have an expansion in powers of the strong coupling $\alpha_s$. The sum extends over all possible partonic channels contributing at a given order in this expansion. At leading order ($\sim\alpha_s^0$) only the channels $(ij)=(q\bar q), (\bar q q)$ contribute, while at NLO ($\sim\alpha_s$) one must include $(ij)=(q\bar q), (\bar q q), (qg), (gq), (\bar q g), (g\bar q)$ in the sum. 

It will be useful for our purposes to introduce the ratios
\begin{equation}\label{zdef}
   \tau = \frac{M^2}{s} \,, \qquad
   z = \frac{M^2}{\hat s} = \frac{\tau}{x_1 x_2} \,,
\end{equation}
where $\hat s=x_1 x_2 s$ is the center-of-mass energy squared of the partonic subprocess that creates the lepton pair. This determines the maximum energy transferred to the leptons and the maximum invariant mass $M$ they can have. In \cite{Anastasiou:2003yy} the coefficient functions $\widetilde C_{ij}$ are expressed in terms of the variable $z$ and a second quantity
\begin{equation}
   y = \frac{\frac{x_1}{x_2}\,e^{-2Y}-z}%
            {(1-z)(1+\frac{x_1}{x_2}\,e^{-2Y})} \,.
\end{equation}
These variables take values on the intervals $0\le y\le 1$ and $\tau\le z\le 1$ subject to the condition that the parton momentum fractions
\begin{equation}\label{xi}
   x_1 = \sqrt{\frac{\tau}{z}\,\frac{1-(1-y)(1-z)}{1-y(1-z)}}\,e^Y \,, 
    \qquad
   x_2 = \sqrt{\frac{\tau}{z}\,\frac{1-y(1-z)}{1-(1-y)(1-z)}}\,e^{-Y} 
\end{equation}
do not exceed 1. The allowed range for the rapidity is such that $2|Y|\le\ln(1/\tau)$. We then define new kernels via
\begin{equation}
   \widetilde C_{ij}(x_1,x_2,s,M,\muf)
   = \left| \frac{dz\,dy}{dx_1\,dx_2} \right|\,
   \frac{C_{ij}(z,y,M,\muf)}{[1-y(1-z)][1-(1-y)(1-z)]} \,.
\end{equation}
At NLO the explicit results for these functions can be written in the form (with $\alpha_s\equiv\alpha_s(\muf)$ and $e_q$ denoting the electric charges of the quarks in units of $e$) \cite{Anastasiou:2003yy}
\begin{eqnarray}\label{Cijres}
   \frac{C_{q\bar q}}{e_q^2}
   &=& \delta(1-z)\,\frac{\delta(y)+\delta(1-y)}{2} \left[ 1
    + \frac{C_F\alpha_s}{\pi} \left( \frac32\ln\frac{M^2}{\muf^2}
    + \frac{2\pi^2}{3} - 4 \right) \right] \nonumber\\
   &&\mbox{}+ \frac{C_F\alpha_s}{\pi}\,\Bigg\{
    \frac{\delta(y) + \delta(1-y)}{2}\,\bigg[ 
    (1+z^2) \left[ \frac{1}{1-z} \ln\frac{M^2(1-z)^2}{\muf^2 z}
     \right]_+ + 1 - z \bigg] \nonumber\\
   &&\quad \mbox{}+ \frac12 
    \left[ 1 + \frac{(1-z)^2}{z}\,y(1-y) \right]
    \left[ \frac{1+z^2}{1-z} \left( \left[ \frac{1}{y} \right]_+ 
    + \left[ \frac{1}{1-y} \right]_+ \right) - 2(1-z) \right]
    \Bigg\} \,, \nonumber\\
   \frac{C_{qg}}{e_q^2}
   &=& \frac{T_F\alpha_s}{2\pi}\,\Bigg\{ \delta(y) \left[
    \left( z^2 + (1-z)^2 \right) \ln\frac{M^2(1-z)^2}{\muf^2 z}
     + 2z(1-z) \right] \nonumber\\
   &&\quad \mbox{}+ \left[ 1 + \frac{(1-z)^2}{z}\,y(1-y) \right]
    \left[ \left( z^2 + (1-z)^2 \right) \left[ \frac{1}{y} \right]_+
    + 2z(1-z) + (1-z)^2 y \right] \Bigg\} \,. \quad
\end{eqnarray}
The $\muf$-dependent terms can be derived from the fact that the cross section in (\ref{sig}) is scale independent, while the PDFs obey the DGLAP evolution equations \cite{Gribov:1972ri,Dokshitzer:1977sg,Altarelli:1977zs}. The remaining functions follow from the symmetry relations
\begin{equation}
   C_{\bar q q} = C_{q\bar q} \,, \quad
   C_{\bar q g} = C_{qg} \,, \quad
   C_{gq} = C_{g\bar q} = C_{qg}|_{y\to 1-y} \,.
\end{equation}
The hard-scattering kernels at NNLO have been calculated in \cite{Anastasiou:2003yy,Anastasiou:2003ds} and are available in the form of a computer program \cite{Vrap}.

The explicit expressions for the coefficient functions given above contain terms that are singular in the ``partonic threshold region" $z\to 1$, in which the center-of-mass energy of the parton subprocess is just large enough to create a lepton pair with invariant mass $M$. Indeed, the arguments of the logarithms in (\ref{Cijres}) suggest the relevance of two mass scales: a ``hard" scale $\muh\sim M$, and a ``soft" scale $\mui\sim M(1-z)/\sqrt{z}=\sqrt{\hat s}\,(1-z)$. Physically, the hard scale is set by the invariant mass of the lepton pair, while the soft scale is of the order of the energy of the remnant jet $X$ produced in the collision. In the region of parton kinematics where $z\to 1$ these scales are separated, $\muh\gg\mui$, in which case the coefficient functions contain large logarithms irrespective of the choice of the factorization scale $\muf$. Threshold resummation for the Drell-Yan cross section \cite{Sterman:1986aj,Catani:1989ne,Magnea:1990qg,Korchemsky:1993uz} aims at resumming these logarithms to all orders in perturbation theory.

Let us return to the structure of the relations (\ref{Cijres}) and identify the leading singular terms in the partonic threshold region. They are contained in $C_{q\bar q}$ and up to NLO multiply $\delta$-functions in the variable $y$. Beyond NLO some of the leading singular terms in the expressions obtained in \cite{Anastasiou:2003yy} multiply nontrivial functions of $y$, but since the $y$-dependence of the parton variables $x_1$ and $x_2$ in (\ref{xi}) is subleading in the $z\to 1$ limit one can always rearrange the expressions in such a way that the leading singular terms multiply $\delta$-functions in $y$. Explicitly, we then obtain
\begin{equation}\label{Cdecomp}
   C_{q\bar q}
   = \frac{\delta(y)+\delta(1-y)}{2}\,e_q^2\,C(z,M,\muf) 
   + C_{q\bar q}^{\rm subl} \,,
\end{equation}
where
\begin{eqnarray}\label{LLterms}
   C(z,M,\muf)
   &=& \delta(1-z)
    + \frac{C_F\alpha_s}{\pi} \left\{ 
     \delta(1-z) \left( \frac32\,L + \frac{2\pi^2}{3} - 4 \right)
     + 2 \left[ \frac{L_z}{1-z} \right]_+ \right\} \nonumber\\
   &&\mbox{}+ C_F \left( \frac{\alpha_s}{\pi} \right)^2 
    \Big[ C_F P_F(z) + C_A P_A(z) + T_F n_f P_f(z) \Big] \,,
\end{eqnarray}
and we have defined $L=\ln(M^2/\muf^2)$ and $L_z=\ln[M^2(1-z)^2/\muf^2 z]$. The terms in the first line can be readily read off from (\ref{Cijres}). The two-loop coefficients $P_i(z)$ are given in Appendix~\ref{app:a}. Note that the factor $z$ in the argument of the logarithm $L_z$ could be set to 1 at leading order, but it is correctly reproduced by our resummation formula below and so we will keep it. The goal of this paper is to derive a formalism that resums these leading terms to all orders in perturbation theory. 

\newpage
Upon performing the integration over $y$, the leading singular terms in (\ref{LLterms}) give rise to the following contribution to the cross section:
\begin{eqnarray}\label{final}
   \frac{d^2\sigma^{\rm thresh}}{dM^2 dY}
   &=& \frac{4\pi\alpha^2}{3N_c M^2 s} \sum_q e_q^2
    \int\frac{dz}{z}\,C(z,M,\muf) \\
   &&\hspace{-2cm}\times \left[
    \frac{f_{q/N_1}(\sqrt{\tau}\,e^Y,\muf)\,
          f_{\bar q/N_2}(\sqrt{\tau}/z\,e^{-Y},\muf)
          + f_{q/N_1}(\sqrt{\tau}/z\,e^Y,\muf)\,
          f_{\bar q/N_2}(\sqrt{\tau}\,e^{-Y},\muf)}{2}
    + (q\leftrightarrow\bar q) \right] . \nonumber
\end{eqnarray}
The lower limit of the $z$ integral is $\sqrt{\tau}\,e^{\mp Y}$, as appropriate for the two terms. At tree level, we recover the parton-model result
\begin{equation}
   \frac{d^2\sigma}{dM^2 dY}
   = \frac{4\pi\alpha^2}{3N_c M^2 s} \sum_q e_q^2 \left[
    f_{q/N_1}(\sqrt{\tau}\,e^Y,\muf)\,
    f_{\bar q/N_2}(\sqrt{\tau}\,e^{-Y},\muf)
    + (q\leftrightarrow\bar q) \right] .
\end{equation}
We also note that integrating over rapidity, we obtain for the leading singular terms in the single-differential cross section
\begin{equation}\label{dsigdM2}
   \frac{d\sigma^{\rm thresh}}{dM^2}
   = \frac{4\pi\alpha^2}{3N_c M^2 s} \sum_q e_q^2 
   \int\frac{dx_1}{x_1} \frac{dx_2}{x_2}\,C(z,M,\muf)
   \left[ f_{q/N_1}(x_1,\muf)\,f_{\bar q/N_2}(x_2,\muf)
    + (q\leftrightarrow\bar q) \right] ,
\end{equation}
where $z=\tau/(x_1 x_2)$, and the integration is restricted to the region where $x_1 x_2\ge\tau$. This result can be rewritten in the more convenient form
\begin{equation}\label{dsigdM2alt}
   \frac{d\sigma^{\rm thresh}}{dM^2}
   = \frac{4\pi\alpha^2}{3N_c M^2 s} \int_\tau^1\!\frac{dz}{z}\,
   C(z,M,\muf)\,\ff(\tau/z,\muf) \,,
\end{equation}
where
\begin{equation}\label{ffdef}
   \ff(y,\muf) = \sum_q e_q^2 \int_y^1\!\frac{dx}{x}
   \left[ f_{q/N_1}(x,\muf)\,f_{\bar q/N_2}(y/x,\muf)
    + (q\leftrightarrow\bar q) \right]
\end{equation}
denotes the Mellin convolution of the PDFs.

In the following section we will derive a factorization formula for the coefficient $C(z,M^2,\muf)$ using methods of effective field theory. The result is
\begin{equation}\label{factform}
   C(z,M,\muf) = H(M,\muf)\,S(\sqrt{\hat s}\,(1-z),\muf) \,, 
\end{equation}
where $H$ and $S$ will be referred to as hard and soft functions, respectively, and will be defined in terms of Wilson coefficients of operators in SCET. The calculation of the components $H$ and $S$ at any order in perturbation theory is much simpler than the calculation of the Drell-Yan cross section at the same order. Eq.~(\ref{factform}) thus provides an approximation to the cross section that requires a minimal amount of calculational work. The all-order resummation of the partonic threshold logarithms is then achieved by solving RG equations.

It must be emphasized at this point that the variable $z$ is not set by external kinematics, but instead is integrated over the interval between $\tau=M^2/s$ and 1. It is therefore necessary to specify under which conditions the partonic threshold region requires special attention. For instance, also in inclusive processes such as $e^- e^-\to\mbox{hadrons}$ there are kinematical situations where scales much smaller than the center-of-mass energy are important, such as the emission of a soft gluon into the final state. It is a well-known fact, however, that upon integration over the entire phase space the perturbation series is insensitive to scales lower than $\sqrt{s}$ at leading power.

There are two limits in which the threshold contributions are parametrically enhanced. First, if the invariant mass of the lepton pair is near the kinematic limit set by the total center-of-mass energy of the hadron-hadron collision, then $\tau\approx 1$ and hence $z\ge\tau$ is always near 1. Threshold resummation is necessary in this case and proceeds in close analogy to the resummation of threshold logarithms for the DIS structure functions for $x\to 1$ \cite{Sterman:1986aj,Catani:1989ne,Magnea:1990qg,Korchemsky:1993uz}. In practice, the region $\tau\approx 1$ is irrelevant for phenomenology, since the strong suppression of the PDFs near the endpoint implies a very low parton luminosity in this case. 

\begin{figure}
\begin{center}
\psfrag{la}[]{}
\psfrag{x}[]{$y$}
\psfrag{y}[b]{$\ff(y,\muf)$}
\includegraphics[width=0.5\textwidth]{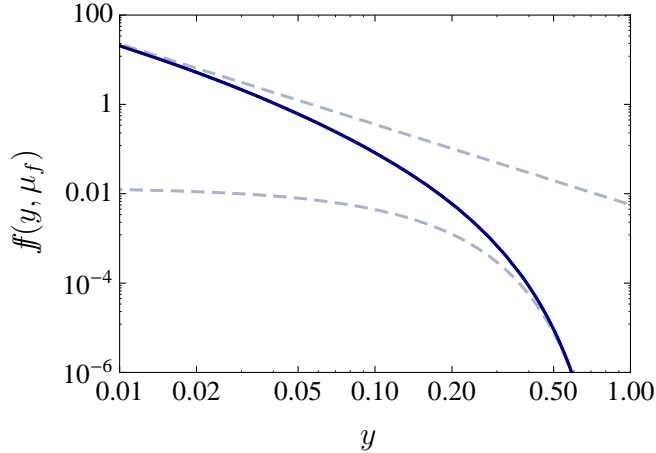} 
\end{center}
\vspace{-0.5cm}
\caption{
\label{fig:ffdrop}
Fall-off of the parton luminosity function $\ff(y,\muf)$ for $\muf=8$\,GeV. The dashed lines show the asymptotic behavior for small and large $y$.}
\end{figure}

A second way in which the threshold contributions can be enhanced arises dynamically, if the weight function multiplying the hard-scattering kernel under the $z$-integral is steeply falling with $(1-z)$ \cite{Appell:1988ie, Catani:1998tm}. In this case threshold resummation can be justified even if $\tau$ is much less than 1. In practice, such a behavior has to result from the fall-off of the parton densities with increasing $x$. Consider for simplicity the total cross section $d\sigma/dM^2$, for which the relevant combination of PDFs is given by the function $\ff(y,\muf)$ in (\ref{ffdef}). Figure~\ref{fig:ffdrop} shows that this function is indeed very steeply falling with $y$. Taking $\muf=8$\,GeV, one finds that $\ff(y,\muf)\propto y^a$ for $y\to 0$ and $\ff(y,\muf)\propto(1-y)^b$ for $y\to 1$, where $a\approx -1.8$ and $b\approx 11$. The figure shows that the first form reasonably well describes the behavior for $y<0.05$, while the second form holds for $y>0.3$. Using these asymptotic forms for the parton luminosity function, we find that for $\tau<0.05$
\begin{equation}
   \frac{d\sigma^{\rm thresh}}{dM^2}
   \approx \frac{4\pi\alpha^2}{3N_c M^2 s}\,\ff(\tau,\muf)
   \int_\tau^1\!\frac{dz}{z}\,z^{-a}\,C(z,M,\muf) \,,
\end{equation}
while for $\tau>0.3$
\begin{equation}
   \frac{d\sigma^{\rm thresh}}{dM^2}
   \approx \frac{4\pi\alpha^2}{3N_c M^2 s}\,\ff(\tau,\muf)
   \int_\tau^1\!\frac{dz}{z} 
   \left( \frac{1-\tau/z}{1-\tau} \right)^b C(z,M,\muf) \,.
\end{equation}
In the first case the cross section is given by a low-order moment of the hard-scattering kernel $C$, in which case the partonic threshold region $z\to 1$ is not parametrically enhanced. In the second case, on the other hand, in the limit where we treat the exponent $b$ as a large parameter, the $z$ integral receives important contributions only from the region where $(1-z)<(1-\tau)/b$. Even for $\tau$ values not near 1 there is thus a parametric enhancement of the partonic threshold region, which turns the threshold logarithms into logarithms of the exponent $b$. The intermediate range $0.05<\tau<0.3$ is a transition region, in which the dynamical enhancement of the threshold region ceases to be effective as $\tau$ is lowered.

\begin{figure}
\begin{center}
\begin{tabular}{cc}
\psfrag{la}[]{}
\psfrag{x}[]{\small $Y$}
\psfrag{y}[b]{\small $d\sigma/dM\,dY$\,[pb/GeV]}
\includegraphics[width=0.45\textwidth]{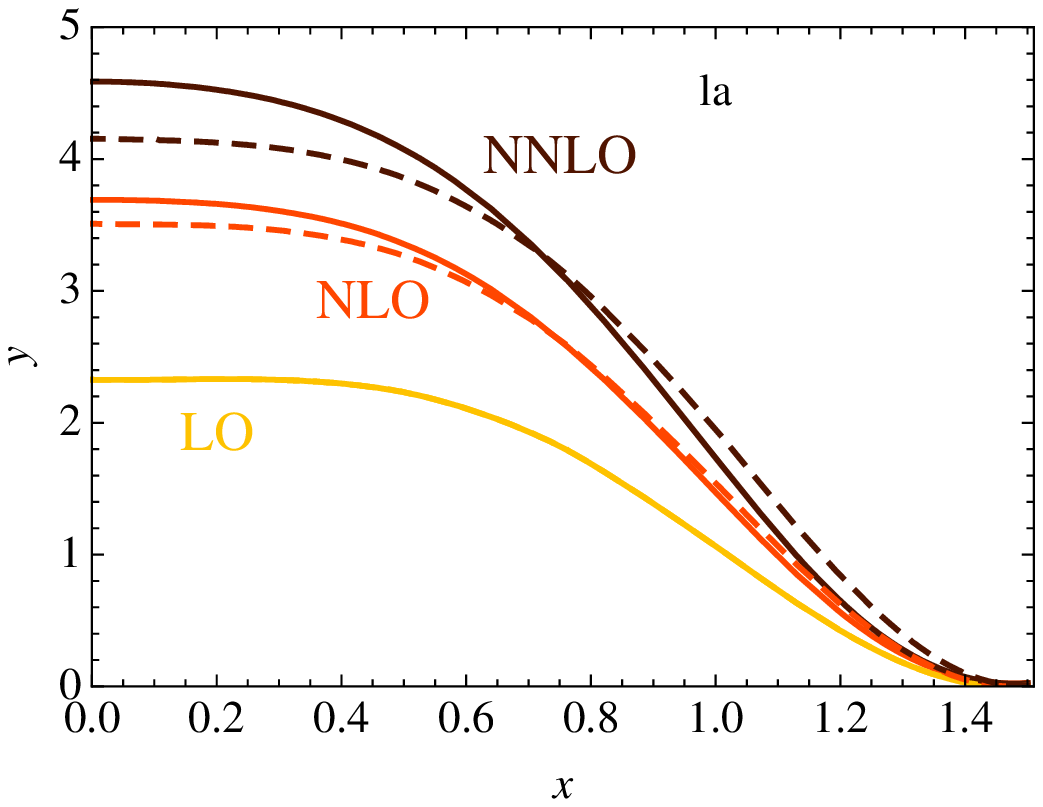} & 
\psfrag{x}[]{\small $Y$}
\psfrag{y}[b]{\small $d\sigma/dM\,dY$\,[pb/GeV]}
\psfrag{la}[]{\footnotesize
 \phantom{abcdefghijkln}\begin{minipage}{3cm} 
 \begin{eqnarray} 
 &&\phantom{a} \nonumber\\[-0.3cm]
 Q &\!=\!& 38.76\,\mbox{GeV} \nonumber\\[-0.25cm]
 M &\!=\!& 8\,\mbox{GeV} \nonumber\\[-0.25cm]
 \muf &\!=\!& M \nonumber
 \end{eqnarray}
 \end{minipage}}
\includegraphics[width=0.45\textwidth]{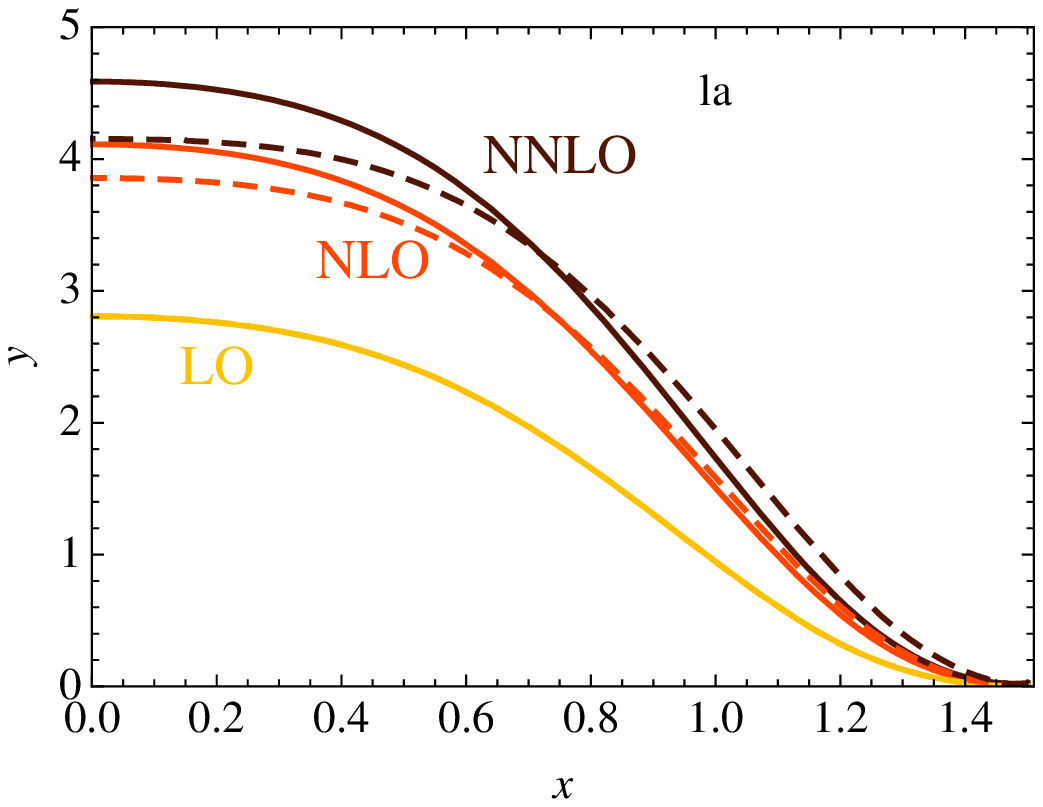}
\end{tabular}
\end{center}
\vspace{-0.5cm}
\caption{
\label{fig:fixedVSresummed}
Comparison of the complete fixed-order results (solid lines) and the contributions arising from the leading singular terms (dashed lines) to the Drell-Yan rapidity distribution at different orders in perturbation theory. On the left we use the PDF sets MRST01LO, MRST04NLO, and MRST04NNLO, as appropriate to the order of the calculation; in the right plot MRST04NNLO is used throughout.}
\end{figure}

While, as we have just discussed, the dominance of the partonic threshold region cannot be justified parametrically for $\tau<0.05$, it nevertheless appears that even in this case the Drell-Yan cross section and rapidity distribution are well approximated by keeping only the leading singular terms in the hard-scattering kernel (\ref{Cdecomp}). The reason is an inherent property of the hard-scattering kernel, which appears to receive the largest radiative corrections from the region of phase space corresponding to Born kinematics. In other words, the effects of hard real emissions appear to be suppressed compared with virtual corrections and soft emissions. To illustrate this point, we show in Figure~\ref{fig:fixedVSresummed} the full fixed-order predictions for the Drell-Yan rapidity distribution at $M=8$\,GeV and $\sqrt{s}=38.76$\,GeV and compare them to the results obtained by keeping only the threshold terms in the hard-scattering kernel. We use the MRST sets of PDFs compiled in \cite{Martin:2004ir}. Throughout our work we use the three-loop running coupling $\alpha_s(\mu)$ normalized to $\alpha_s(M_Z)=0.1167$, and we take $n_f=5$ for the number of light quark flavors. In the right plot we use the same NNLO parton densities for all curves, so that the size of the different perturbative corrections to the hard-scattering kernel can be read off directly. Since our main focus in this work is on the behavior of the perturbative expansion of the hard-scattering kernel, we will from now on always use the set MRST04NNLO for the PDFs. Even though $\tau\approx 0.04$ is very small in this example, we observe that at central rapidity about 80\% of the NLO correction and 63\% of the NNLO correction arise from the leading singular terms. For the total cross section $d\sigma/dM^2$ about 93\% of the NLO correction and 97\% of the NNLO correction are accounted for by the leading singular terms. If we lower the factorization scale to $\muf=M/2$, then the leading singular terms come even closer to reproducing the fixed-order results.

\section{Derivation of the factorization formula}
\label{sec:SCET}

The factorization theorem (\ref{factform}) for the leading singular terms in the hard-scattering kernel $C$ has been established a long time ago \cite{Sterman:1986aj,Catani:1989ne,Magnea:1990qg,Korchemsky:1993uz}. In particular, it has been understood that the soft function $S$ can be represented as the vacuum expectation value of a certain Wilson loop of soft gluon fields. Nevertheless we find it useful to rederive this formula using methods of effective field theory. The advantage of this approach is that we will relate the hard and soft functions, $H$ and $S$, to Wilson coefficients of operators in the effective theory, which obey certain RG equations. Solving these equations we accomplish threshold resummation directly in momentum space.

In the framework of SCET the standard factorization formula (\ref{sig}) for the Drell-Yan cross section has been discussed in \cite{Bauer:2002nz}. However, the issue of Glauber gluons \cite{Bodwin:1981fv, Collins:1981tt} has not yet been addressed in the effective theory. Glauber-gluon interactions have presented an important complication in the factorization analysis of the Drell-Yan process \cite{Collins:1985ue,Collins:1988ig}, and it would be worthwhile to rederive their cancellation using SCET. As discussed below, Glauber gluons do not affect the factorization of the hard-scattering kernel $C(z,M,\muf)$ in the threshold region. 

The threshold resummation for the total cross section $d\sigma/dM^2$ has been performed in \cite{Idilbi:2006dg} using an effective-theory variant of the traditional Mellin moment-space approach. In the present paper we go beyond these works by presenting a derivation of the factorization formula (\ref{factform}) in SCET and performing the threshold resummation directly in $z$ space. This will provide us with the tools to discuss the relevance of resummation away from the true threshold, i.e., for $M^2$ much smaller than $s$.

We begin with the standard formula for the Drell-Yan cross section,
\begin{equation}\label{textbook}
   d\sigma = \frac{4\pi\alpha^2}{3s q^2}\,\frac{d^4q}{(2\pi)^4}
   \int d^4x\,e^{-iq\cdot x}\,
   \langle N_1(p_1) N_2(p_2)|(-g_{\mu\nu}) J^{\mu\dagger}(x)
   J^\nu(0)|N_1(p_1) N_2(p_2)\rangle \,,
\end{equation}
where $J^\mu=\sum_q e_q\,\bar q\gamma^\mu q$ is the electromagnetic current. To derive the factorization theorem, we match the product of currents onto operators in SCET. The matching proceeds in two steps, and the corresponding Wilson coefficients are the hard function $H$ and the soft function $S$, respectively. The remaining effective-theory matrix element can be identified with the PDFs. A similar two-step matching procedure has been used in many SCET applications. In particular, the factorization theorem for DIS in the region $x\to 1$ was derived in SCET by proceeding this way \cite{Becher:2006mr}.

In the present paper we are solely interested in the factorization of the hard-scattering coefficient $C(z,M,\muf)$. For this purpose, we can simplify the effective-theory analysis by considering (\ref{textbook}) with partonic instead of hadronic matrix elements. These partonic matrix elements correspond directly to the hard-scattering coefficients, and the hadronic cross sections are then obtained after convoluting the results with the PDFs. An effective-theory analysis of the hadronic matrix elements would be more complicated because of their sensitivity to nonperturbative physics governed by the scale $\Lambda_{\rm QCD}$, which does not enter the hard-scattering kernels. 

Note that the product of current operators in the matrix element in (\ref{textbook}) is not time-ordered. This is in contrast to the case of DIS, where the decay rate can be obtained from the discontinuity of a time-ordered product of currents. It is not possible to rewrite (\ref{textbook}) in the same way, because the imaginary part of the time-ordered product also gets contributions from virtual corrections to $N_1+N_2\to X$, where $X$ is a purely hadronic final state. A path integral framework to analyze operator products that are not time ordered is the Keldysh formalism \cite{Schwinger:1960qe,Keldysh:1964ud}, which we review in Appendix~\ref{sec:keldysh}. In our discussion below, the operator ordering does not lead to any complications, but it implies that the soft function is not given by Feynman diagrams, but appropriate cuts of such diagrams, as explained in \cite{Belitsky:1998tc}.

\subsection{Derivation in SCET}

\begin{figure}
\begin{center}
\psfrag{qp}{$\bar q$}
\psfrag{qm}{$q$}
\psfrag{lm}[b]{$\phantom{!!!!}\ell^+$}
\psfrag{lp}[b]{$\phantom{!!}\ell^-$}
\psfrag{H}[b]{\large $\,\,\,H$}
\includegraphics[width=0.32\textwidth]{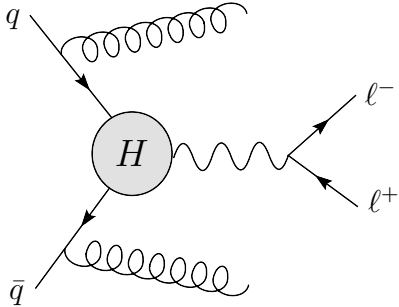}
\end{center}
\vspace{-0.5cm}
\caption{\label{fig:cartoon}
Soft gluon emissions from the initial-state partons in Drell-Yan production.}
\end{figure}

Before entering the technicalities of the discussion of factorization in SCET, it is useful to understand the physics behind the factorization theorem (\ref{factform}) by means of a simple, intuitive argument. Consider the special kinematics of Drell-Yan production in the (partonic) threshold region, as illustrated in Figure~\ref{fig:cartoon}. Because the partonic center-of-mass energy is just above the invariant mass of the produced Drell-Yan pair, only soft emissions from the initial-state partons are allowed. As is well known, these soft emissions are described by eikonal interactions and exponentiate into Wilson lines. Furthermore, to leading power they leave the incoming partons on the mass shell, so that the production of the Drell-Yan pair is described by an on-shell quark form factor.

We will now formalize this argument using the language of effective field theory. The effective theory is constructed by introducing fields for the momentum regions that contribute to the matrix elements in the given kinematics. As stressed above, for our purposes it is sufficient to analyze partonic matrix elements. To distinguish the different momentum modes, we introduce the light-cone decomposition
\begin{equation}
   p^\mu = (n\cdot p)\,\frac{\bar n^\mu}{2}
    + (\bar n\cdot p)\,\frac{n^\mu}{2} + p_\perp^\mu
   \equiv p_+^\mu + p_-^\mu + p_\perp^\mu \,,
\end{equation}
where $n^\mu=(1,0,0,1)$ and $\bar n^\mu=(1,0,0,-1)$ are two light-like reference vectors ($n\cdot\bar n=2$) in the directions of the colliding partons. We denote the small expansion parameter by $\epsilon=(1-z)\ll 1$ and quote the components $(p_+,p_-,p_\perp)$ of parton momenta. The relevant momentum regions are
\begin{eqnarray}
   \mbox{hard:} \quad\,\,
    p_{h} &\sim& \sqrt{\hat s}\,(1,1,1) \,, \nonumber\\
   \mbox{hard-collinear:} \quad
    p_{hc} &\sim& \sqrt{\hat s}\,(\epsilon,1,\sqrt{\epsilon}) \,,
    \nonumber\\
   \mbox{anti-hard-collinear:} \quad
    p_{\overline{hc}} &\sim& \sqrt{\hat s}\,
    (1,\epsilon,\sqrt{\epsilon}) \,, \nonumber\\
   \mbox{soft:} \quad\,\,
    p_s &\sim& \sqrt{\hat s}\,(\epsilon,\epsilon,\epsilon) \,,
    \nonumber
\end{eqnarray} 
where the large momentum scale in the process is set by the partonic center-of-mass energy $\sqrt{\hat s}$. In the first matching step, the contributions from the hard region are absorbed into Wilson coefficients, while the remaining contributions are represented by fields in the effective theory. The version of the effective theory with fields scaling in the above way is often called SCET$_{\rm I}$. Note that, while in many applications the soft scale is associated with nonperturbative physics, in our case $p_s^2\sim\hat s(1-z)^2$ is of order the mass of the hadronic final-state jet and assumed to be in the perturbative domain. As explained above, the kinematics of the Drell-Yan process near threshold is such that the hadronic final state is made up of soft partons; the virtuality of the hard-collinear partons is parametrically larger than the jet mass: $p_{hc}^2\sim p_{\overline{hc}}^2\sim\hat s(1-z)\gg M_X^2\sim\hat s(1-z)^2$.

The potential contribution from Glauber gluons \cite{Bodwin:1981fv, Collins:1981tt}
presents a challenge when deriving the standard factorization theorem of the Drell-Yan cross section into PDFs convoluted with hard partonic cross sections \cite{Collins:1985ue,Collins:1988ig}. However, the Glauber region does not present an additional difficulty for the factorization of the hard-scattering coefficient $C(z, M,\mu _f)$ in the threshold region, which we will derive below. The hard-scattering coefficient is given by the on-shell $q \bar{q} \to l^+l ^-  + X$ cross section, so that for its factorization only Glauber exchanges involving active quarks need to be studied. The momenta of the Glauber gluons relevant for the threshold region scale as $p\sim \sqrt{\hat s} (\epsilon, \epsilon , \sqrt{\epsilon} ) $, and these gluons are thus not allowed in the hadronic final state, which is assumed to have $M_X^2 \sim \epsilon^2 \hat s$. Writing down the loop integrals for Glauber exchanges between the two inital-state quarks, one finds that they are scaleless and vanish in dimensional regularization. In more physical terms, this means that the Glauber contribution is accounted for already by the standard momentum regions introduced above. 

In order not to clutter the notation, we consider for the moment a single quark flavor with charge $e_q=1$. We thus match the current $J^\mu=\bar\psi\gamma^\mu\psi$ onto an effective current operator in SCET containing a hard-collinear quark and an anti-hard-collinear anti-quark \cite{Becher:2006mr,Bauer:2002nz,Becher:2003kh,Manohar:2003vb}. At leading power in $\epsilon$, this yields 
\begin{equation}
   J^\mu(0)\to \int\!ds\,dt\,\widetilde C_V(s,t,\muf)\,
   \big( \bar\xi_{\overline{hc}} W_{\overline{hc}} \big)(sn)\,
   \gamma_\perp^\mu\,\big( W_{hc}^\dagger\xi_{hc} \big)(t\bar n) \,,
\end{equation}
where $W_{hc}$ and $W_{\bar hc}$ are the usual hard-collinear (along the $\bar n$-direction) and anti-hard-collinear (along the $n$-direction) Wilson lines of SCET. Note that the matching relation for the (transverse) vector current does not include a two-gluon operator at leading power in the effective theory. Such an operator would arise, however, in the corresponding matching relation for a scalar current relevant for Higgs production. Mixed quark-gluon operators corresponding to the $(ij)= (gq)$, $(qg)$, $(\bar q g)$, and $(g\bar q)$ scattering channels are power suppressed. Accordingly, the corresponding hard-scattering kernels $C_{ij}$ do not contain terms that are singular in the limit $z\to 1$, as is explicitly seen in (\ref{Cijres}). The Fourier transform of the position-space Wilson coefficient
\begin{equation}
   C_V(-\bar n\cdot p_1\,n\cdot p_2,\muf)
   = \int\!ds\,dt\,\widetilde C_V(s,t,\muf)\,
   e^{-is\bar n\cdot p_1-itn\cdot p_2}
\end{equation}
is a function of the product of the large light-cone momentum components carried by the quark fields. In our case, $\bar n\cdot p_1\,n\cdot p_2=q^2$ is equal to the hard scale set by the mass of the Drell-Yan pair. As a result the Wilson coefficient is evaluated at the time-like momentum transfer carried by the current, $C_V(-q^2-i\epsilon,\muf)$. The $i\epsilon$ prescription is required since this function has a branch cut along the positive real $q^2$ axis. Note that the same Wilson coefficient $C_V$ appears in DIS, but evaluated at space-like momentum transfer \cite{Becher:2006mr,Idilbi:2006dg,Manohar:2003vb}. The coefficient $C_V$ can be determined by on-shell matching; indeed, it is simply given by the on-shell massless form factor in QCD \cite{Becher:2006nr}. The Drell-Yan cross section involves the current squared, so that the hard function in (\ref{factform}) is given by
\begin{equation}\label{Hdef}
   H(M,\muf) = |C_V(-M^2-i\epsilon,\muf)|^2 \,.
\end{equation}
The expression for the coefficient $C_V$ up to $O(\alpha_s^2)$ was derived in \cite{Becher:2006mr,Idilbi:2006dg} and can be found in Appendix~\ref{Appendix:RGfunctions}.

At leading power in $\epsilon$, only the $n\cdot A_s$ component of the soft gluon field couples to the hard-collinear fields. These eikonal interactions can be represented by Wilson lines. In the effective theory this is achieved by redefining the hard-collinear fields as  
\cite{Bauer:2001yt,Becher:2003qh}
\begin{equation}\label{decoupl}
   \xi_{hc}(x)\to S_n(x_-)\,\xi_{hc}^{(0)}(x) \,, \qquad
   A_{hc}^\mu(x)\to S_n(x_-)\,A_{hc}^{\mu(0)}(x)\,S_n^\dagger(x_-) \,,
\end{equation}
which implies $(W_{hc}^\dagger\xi_{hc})(x)\to S_n(x_-)\,(W_{hc}^\dagger\xi_{hc})^{(0)}(x)$. Here
\begin{equation}\label{eq:Sn}
   S_n(x) = {\rm\bf P}\,\exp\left(
   ig\int_{-\infty}^0\!ds\,n\cdot A_{s}(x+sn) \right)
\end{equation}
is a soft Wilson line along the $n$ light-cone. The same redefinition, but with $n$ and $\bar n$ interchanged, decouples the soft gluon field also from the anti-hard-collinear fields. As a result, the current operator 
\begin{equation}
    J^\mu(0)\to 
    \big( \bar\xi_{\overline{hc}} W_{\overline{hc}} \big)^{(0)}(s n)
    \,\gamma_\perp^\mu\,
    \big( S_{\bar n}^\dagger S_n \big)(0)\, 
    \big( W_{hc}^\dagger\xi_{hc} \big)^{(0)}(t\bar n)
\end{equation}
splits into three parts, which no longer interact with each other. In the same way, the matrix element for the Drell-Yan process factorizes in the form
\begin{eqnarray}\label{eq:factor}
   && \langle N_1(p_1) N_2(p_2)|(-g_{\mu\nu})
    J^{\mu\dagger}(x) J^\nu(0)|N_1(p_1) N_2(p_2)\rangle \nonumber\\
   &\to& \frac{1}{N_c} \left| C_V(-M^2-i\epsilon,\muf) \right|^2\,
    {\hat W}_{\rm DY}(x,\muf)\,
    \langle N_1|(\bar\xi_{hc} W_{hc})^{(0)}(x_+)\,
    \frac{\nbslash}{2}\,
    (W_{hc}^\dagger\xi_{hc})^{(0)}(0)|N_1\rangle \nonumber\\
   &&\times \langle N_2|
    (\bar\xi_{\overline{hc}} W_{\overline{hc}})^{(0)}(0)\,
    \frac{\nslash}{2}\,
    (W_{\overline{hc}}^\dagger\xi_{\overline{hc}})^{(0)}(x_-)
    |N_2\rangle \,.
\end{eqnarray}
To obtain this expression, we have Fierz rearranged the fermion fields and have averaged over the color of the external states. We have simplified the Dirac algebra making use of the projection properties $\nslash\,\xi_{hc}=0$ and $\nbslash\,\xi_{hc}=\xi_{hc}$ of the hard-collinear fermion fields (and likewise $\nbslash\,\xi_{\overline hc}=0$ and $\nslash\,\xi_{\overline hc}=\xi_{\overline hc}$ for the anti-hard-collinear fields). Also, we have neglected the power-suppressed dependence of the hard-collinear matrix element on $x_-$ and $x_\perp$ ($x_+$ and $x_\perp$ for the anti-hard-collinear matrix element), using the fact that up to power corrections the incoming partons fly along the beam axis. In more technical terms, we have multi-pole expanded the corresponding fields to leading power \cite{Beneke:2002ph,Beneke:2002ni}. The soft matrix element ${\hat W}_{\rm DY}(x,\muf)$ (not to be confused with the hard-collinear Wilson lines) is a closed Wilson loop formed from the product of the soft Wilson lines in the two currents,
\begin{equation}\label{softmatrixelement}
   {\hat W}_{\rm DY}(x,\muf) = \frac{1}{N_c}\,
   \langle 0|\,\mbox{Tr}\,{\bf \bar T}
   \big[ S_{n}^\dagger(x) S_{\bar n}(x) \big]\,
   {\bf T} \big[ S_{\bar n}^\dagger(0) S_n(0) \big]|0 \rangle \,,
\end{equation}
where the trace is over color indices, and the operator ${\bf\bar T}$ makes explicit that the Wilson lines in the complex conjugate current $J^{\mu\dagger}$ are anti-time-ordered. A detailed discussion of the operator ordering is given in Appendix~\ref{sec:keldysh}. It will become evident in Section~\ref{sec:kin} that the function ${\hat W}_{\rm DY}(x,\muf)$ is closely related to the soft function for the Drell-Yan process. 

In the second matching step we lower the renormalization scale below the soft scale and integrate out the soft fields. Because the soft scale is in the short-distance domain, this simply amounts to a perturbative calculation of the Wilson loop (\ref{softmatrixelement}).
 
To make contact with the standard treatment, where the factorization theorem is derived within perturbative QCD, we note that after the decoupling transformation (\ref{decoupl}) the leading-power Lagrangian in each of the three sectors of the effective theory is completely equivalent to the QCD Lagrangian. We can thus equally well evaluate the soft function with QCD Wilson lines instead of soft Wilson lines. The same is true for the hard-collinear matrix elements, where one can replace
\begin{equation}\label{collinearmatrixelemt}
   \langle N_1|(\bar\xi_{hc} W_{hc})^{(0)}(x_+)\,\frac{\nbslash}{2}\,
   (W_{hc}^\dagger\xi_{hc})^{(0)}(0)|N_1\rangle
   \to \langle N_1|\,\bar\psi(x_+)\,\frac{\nbslash}{2}\,[x_+,0]\,
   \psi(0)|N_1\rangle \,,
\end{equation}
where $\psi(x)$ is the QCD quark field and $[x_+,0]$ a QCD Wilson line of $n\cdot A$ gluons extending from $0$ to $x_+$. 

We have simplified the effective-theory treatment by restricting ourselves to on-shell partonic instead of hadronic matrix elements, which has the advantage that we do not need to discuss momentum regions that scale with powers of $\Lambda_{\rm QCD}$.\footnote{For DIS at $x\to 1$, a full factorization analysis (including messenger fields \cite{Becher:2003qh}) was performed in~\cite{Becher:2006mr}.} 
These momentum regions cannot enter the hard or the soft functions, and given that the Drell-Yan cross section is known to factorize for $\hat s(1-z)\gg\Lambda_{\rm QCD}^2$, we are guaranteed that their presence will not spoil factorization. The restriction to partonic matrix elements can thus safely be dropped. The hadronic matrix elements of the form (\ref{collinearmatrixelemt}) are the usual PDFs \cite{Collins:1981uw}
\begin{equation}\label{phidef}
   f_{q/N}(x,\muf) = \frac{1}{2\pi} \int_{-\infty}^\infty\!dt\,
   e^{-ixt\bar n\cdot p}\,\langle N(p)|\,\bar\psi(t\bar n)\,
   \frac{\nbslash}{2}\,[t\bar n,0]\,\psi(0)\,|N(p)\rangle \,.
\end{equation}

Inserting the factorized form of the matrix element (\ref{eq:factor}) into expression (\ref{textbook}) for the cross section, we obtain \begin{eqnarray}\label{sigma1}
   \frac{d\sigma}{dM^2} 
   &=& \frac{4\pi\alpha^2}{3N_c M^2}\,
    |C_V(-M^2-i\epsilon,\muf)|^2 \sum_q e_q^2 \int dx_1\,dx_2 
    \left[ f_{q/N_1}(x_1,\muf)\,f_{\bar q/N_2}(x_2,\muf)
    + (q\leftrightarrow\bar q) \right] \nonumber\\
   &&\times \int\frac{d^3\vec{q}}{(2\pi)^3 2q^0}\,
    \frac{1}{2\pi} \int d^4x\,
    e^{i(x_1 p_{1-}+x_2 p_{2+}-q)\cdot x}\,\hat W_{\rm DY}(x) \,.
\end{eqnarray}
From now on we will drop the $i\epsilon$ prescription in the argument of $C_V$.
   
\subsection{Kinematic simplifications in the threshold region}
\label{sec:kin}

To proceed, we need to study the kinematics in the partonic threshold region, where $z\to 1$. With the expansion parameter $\epsilon=(1-z)\ll 1$, it is easy to show that in the parton center-of-mass system
\begin{equation}
   |\vec{q}|\le \frac{\sqrt{\hat s}}{2}\,(1-z) = O(\epsilon) 
    \qquad \Rightarrow \qquad
   q^0 = \sqrt{\hat s} + O(\epsilon) \,.
\end{equation}
It follows that $q^0$ is parametrically larger than the spatial components of $q^\mu$. This observation implies 
that the term $q^0$ in the denominator in (\ref{sigma1}) is independent of $\vec{q}$ at leading power, in which case performing the integral over $d^3\vec{q}$ yields $\delta^{(3)}(\vec{x})$. It also follows that the rapidity of the Drell-Yan pair in the parton center-of-mass system vanishes at leading power,
\begin{equation}
   Y_{\rm CMS} = \frac12\ln\frac{q^0+q^3}{q^0-q^3} = O(\epsilon) \,,
\end{equation}
which explains why the rapidity in the laboratory frame, 
$Y=Y_{\rm CMS}+\frac12\ln\frac{x_1}{x_2}$, is determined by the ratio $x_1/x_2$ at leading power in $(1-z)$, as can be seen from (\ref{xi}). 
Finally, we need that
\begin{equation}
   (x_1 p_{1-}+x_2 p_{2+}-q)^0
   = \frac{\sqrt{\hat s}}{2}\,(1-z) + O(\epsilon^2) \,.
\end{equation}

Using these results, we obtain at leading power in the $z\to 1$ limit the expression
\begin{eqnarray}\label{sigma2}
   \frac{d\sigma}{dM^2} &=& \frac{4\pi\alpha^2}{3N_c M^2}\,
    |C_V(-M^2,\muf)|^2 \sum_q e_q^2 \int dx_1\,dx_2
    \left[ f_{q/N_1}(x_1,\muf)\,f_{\bar q/N_2}(x_2,\muf)
    + (q\leftrightarrow\bar q) \right] \nonumber\\
   &&\times \frac{1}{2\sqrt{\hat s}}
    \int\frac{dx^0}{2\pi}\,e^{i\sqrt{\hat s}(1-z)x^0/2}\,
    \hat W_{\rm DY}(x^0,\vec{x}=0,\muf) \,,
\end{eqnarray}
where the integration region over the parton momentum fractions is such that $x_1 x_2\ge\tau$, so that $z=\tau/x_1 x_2\le 1$. In the final step we introduce the Fourier transform of the position-space Wilson loop at time-like separation via
\begin{equation}\label{WDYdef}
   W_{\rm DY}(\omega,\muf) = \int\frac{dx^0}{4\pi}\,
   e^{i\omega x^0/2}\,\hat W_{\rm DY}(x^0,\vec{x}=0,\muf) \,.
\end{equation}
This Wilson loop plays the role of the jet function in DIS in the sense that it describes the properties of the hadronic final state. It has been introduced previously in \cite{Korchemsky:1993uz,Belitsky:1998tc}. The Drell-Yan cross section now takes the form
\begin{eqnarray}\label{sigma3}
   \frac{d\sigma}{dM^2} &=& \frac{4\pi\alpha^2}{3N_c M^2 s}\,
    |C_V(-M^2,\muf)|^2 \sum_q e_q^2
    \int\frac{dx_1}{x_1}\,\frac{dx_2}{x_2}
    \left[ f_{q/N_1}(x_1,\muf)\,f_{\bar q/N_2}(x_2,\muf)
    + (q\leftrightarrow\bar q) \right] \nonumber\\
   &&\times \sqrt{\hat{s}}\,W_{\rm DY}(\sqrt{\hat s}\,(1-z),\muf) \,,
\end{eqnarray}
and from comparison with (\ref{dsigdM2}) and (\ref{factform}) we identify the soft function as
\begin{equation}
   S(\sqrt{\hat s}\,(1-z),\muf) 
   = \sqrt{\hat s}\,W_{\rm DY}(\sqrt{\hat s}\,(1-z),\muf) \,.
\end{equation} 
We should mention that at leading power in $(1-z)$ the argument of the soft function could be simplified as $\sqrt{\hat s}\,(1-z)=M(1-z)/\sqrt{z}\approx M(1-z)$; however, since the exact expressions (\ref{Cijres}) for the hard-scattering kernels at NLO contain the logarithm $L_z=\ln(\hat s(1-z)^2/\muf^2)$, we prefer to keep the argument of the soft function in the form written above.

\section{Momentum-space resummation at large \boldmath $z$\unboldmath}
\label{sec:resummation}

At this point we have identified the two components in the factorized expression (\ref{factform}) for the hard-scattering coefficient $C(z,M,\muf)$ with field-theoretic objects defined in terms of operator matrix elements. The resummation of threshold logarithms arising in the $z\to 1$ region can now be accomplished by solving the RG evolution equations obeyed by these quantities.

\subsection{Evolution of the hard function}

The evolution equation for the hard matching coefficient $C_V$ evaluated at time-like momentum transfer and its solution can be obtained from the corresponding results valid for space-like momentum transfer \cite{Becher:2006nr} by analytic continuation. This leads to
\begin{equation}\label{gammaV}
   \frac{d}{d\ln\mu}\,C_V(-M^2-i\epsilon,\mu)
   = \left[ \Gamma_{\rm cusp}(\alpha_s)
   \left( \ln\frac{M^2}{\mu^2} - i\pi \right)
   + \gamma^V(\alpha_s) \right] C_V(-M^2-i\epsilon,\mu) \,.
\end{equation}
We have reinserted the $i\epsilon$ regulator, which determines the sign of the imaginary part of the anomalous dimension. The appearance of the logarithm and its coefficient, the cusp anomalous dimension $\Gamma_{\rm cusp}$ \cite{Korchemsky:1987wg,Korchemskaya:1992je}, can be explained using arguments presented in \cite{Becher:2003kh}. This term in the evolution equation is associated with Sudakov double logarithms. The remaining term, $\gamma^V$, accounts for single-logarithmic evolution.

The exact solution to (\ref{gammaV}) is
\begin{equation}\label{CVsol}
   C_V(-M^2,\muf) = \exp\left[ 2S(\muh,\muf) - a_{\gamma^V}(\muh,\muf) 
   + i\pi\,a_\Gamma(\muh,\muf) \right]
   \left( \frac{M^2}{\muh^2} \right)^{-a_\Gamma(\muh,\muf)}\,
   C_V(-M^2,\muh) \,,
\end{equation}
where $\muh\sim M$ is a hard matching scale, at which the value of $C_V$ is calculated using fixed-order perturbation theory. Note that the Wilson coefficient at time-like momentum transfer is a complex quantity. The Sudakov exponent $S$ and the exponents $a_n$ are given by \cite{Neubert:2004dd}
\begin{equation}\label{RGEsols}
   S(\nu,\mu) 
   = - \int\limits_{\alpha_s(\nu)}^{\alpha_s(\mu)}\!
    d\alpha\,\frac{\Gamma_{\rm cusp}(\alpha)}{\beta(\alpha)}
    \int\limits_{\alpha_s(\nu)}^\alpha
    \frac{d\alpha'}{\beta(\alpha')} \,, \qquad
   a_\Gamma(\nu,\mu) 
   = - \int\limits_{\alpha_s(\nu)}^{\alpha_s(\mu)}\!
    d\alpha\,\frac{\Gamma_{\rm cusp}(\alpha)}{\beta(\alpha)} \,, 
\end{equation}
and similarly for the function $a_{\gamma^V}$. The perturbative expansions of the anomalous dimensions and the resulting expressions for the evolution functions valid at NNLO in RG-improved perturbation theory are collected in Appendix~\ref{Appendix:RGfunctions}. 

\subsection{Evolution of the soft function}

In order to derive the evolution equation for the soft function it is important to have a consistent definition of the threshold region. To this end we consider the limit $M^2\to s$ in (\ref{sigma3}), which implies $z\to 1$, since $z\ge\tau$. The condition $x_1 x_2\ge\tau$ then implies that $x_{1,2}\to 1$, in which case the DGLAP evolution for the PDFs can be written in the simplified form
\begin{equation}\label{phievol}
   \frac{d}{d\ln\mu}\,f_{q/N}(x,\mu) 
   = 2\gamma^\phi(\alpha_s)\,f_{q/N}(x,\mu) 
    + 2\Gamma_{\rm cusp}(\alpha_s) \int_x^1\!\frac{dz}{z}\,
    \frac{f_{q/N}(x/z,\mu)}{[1-z]_+} + \dots \,,
\end{equation}
which is obtained by expanding the Altarelli-Parisi splitting function as
\begin{equation}\label{Pqq}
   P_{q\leftarrow q}(z)
   = \frac{2\Gamma_{\rm cusp}(\alpha_s)}{[1-z]_+}
   + 2\gamma^\phi(\alpha_s)\,\delta(1-z) + \dots \,.
\end{equation}
This asymptotic form of the splitting function holds to all orders in perturbation theory \cite{Korchemsky:1992xv}. The cusp anomalous dimension and the coefficient $\gamma^\phi$ have been calculated at three-loop order \cite{Moch:2004pa}. They are given in Appendix~\ref{Appendix:RGfunctions}.

Requiring that the Drell-Yan cross section in the threshold region be independent of the arbitrary factorization scale $\muf$, we find that the momentum-space Wilson loop obeys the integro-differential evolution equation
\begin{eqnarray}\label{Jrge}
   \frac{dW_{\rm DY}(\omega,\mu)}{d\ln\mu}
   &=& - \left[ 4\Gamma_{\rm cusp}(\alpha_s)\,\ln\frac{\omega}{\mu}
    + 2\gamma^W(\alpha_s) \right] W_{\rm DY}(\omega,\mu) \nonumber\\
   &&\mbox{}- 4\Gamma_{\rm cusp}(\alpha_s) \int_0^{\omega}\!d\omega'\,
    \frac{W_{\rm DY}(\omega',\mu)-W_{\rm DY}(\omega,\mu)}%
         {\omega-\omega'} \,,
\end{eqnarray} 
where
\begin{equation}\label{gw}
   \gamma^W = 2\gamma^\phi + \gamma^V \,.
\end{equation}
Curiously, this quantity starts only at two-loop order. The analogous relation in the case of DIS is $\gamma^J=\gamma^\phi+\gamma^V$, where the anomalous dimension $\gamma^J$ of the DIS jet function is defined in analogy with $\gamma^W$ in (\ref{Jrge}). Combining the two relations we obtain $2\gamma^J-\gamma^W=\gamma^V$, which is an exact relation that links the anomalous dimensions of the soft function in Drell-Yan production and the DIS jet function to the anomalous dimension of the vector current.

The exact solution to the evolution equation (\ref{Jrge}) can be written in the form \cite{Becher:2006nr,Neubert:2005nt}
\begin{equation}\label{sonice}
   W_{\rm DY}(\omega,\muf)
   = \exp\left[ - 4S(\mui,\muf) + 2 a_{\gamma^W}(\mui,\muf) \right]
   \widetilde s_{\rm DY}(\partial_\eta,\mui)\,
   \frac{1}{\omega}
   \left( \frac{\omega}{\mui} \right)^{2\eta}\,
   \frac{e^{-2\gamma_E\eta}}{\Gamma(2\eta)} \,,
\end{equation}
where $\partial_\eta$ denotes a derivative with respect to an auxiliary parameter $\eta$, which is then identified with $\eta=2a_\Gamma(\mui,\muf)$. This result is well defined for $\eta>0$. The solution for negative $\eta$ is obtained by analytic continuation. For instance, to obtain the result for $-\frac12<\eta<0$ we use the identity
\begin{equation}\label{star}
   \int_0^{\Omega}\!d\omega\,\frac{f(\omega)}{\omega^{1-2\eta}}
   = \int_0^{\Omega}\!d\omega\,\frac{f(\omega)-f(0)}{\omega^{1-2\eta}}
   + \frac{f(0)}{2\eta}\,\Omega^{2\eta} ,
\end{equation}
where $f(\omega)$ is a smooth test function. For $\eta<-\frac12$ additional subtractions are required.

The function $\widetilde s_{\rm DY}$ is obtained from the momentum-space Wilson loop by the Laplace transformation \cite{Becher:2006nr}
\begin{equation}\label{Laplace}
   \widetilde s_{\rm DY}(L,\mui)
   = \int_0^\infty\!d\omega\,e^{-s\omega}\,W_{\rm DY}(\omega,\mui)
    \,, \qquad
   s = \frac{1}{e^{\gamma_E}\mui\,e^{L/2}} \,.
\end{equation}
The position-space Wilson loop on the right-hand side of (\ref{WDYdef}) can be shown to have the functional form $\hat W_{\rm DY}(x^0,\vec{x}=0,\mui)=f(\frac{i}{2}x^0\mui e^{\gamma_E},\alpha_s(\mui))$ \cite{Korchemsky:1993uz,Belitsky:1998tc}, and a straightforward calculation shows that the function $\widetilde s_{\rm DY}(L,\muf)$ can be expressed in terms of $f(t,\alpha_s)$ as
\begin{equation}\label{wonderful}
   \widetilde s_{\rm DY}(L,\mui) = f(e^{-L/2},\alpha_s(\mui)) \,.
\end{equation}
We can then use the explicit two-loop expression for the position-space Wilson loop obtained in \cite{Belitsky:1998tc} to compute the function $\widetilde s_{\rm DY}$ at two-loop order. The result is presented in Appendix~\ref{Appendix:RGfunctions}.

\subsection{Resummation of large logarithms}

We are now in a position to derive the RG-resummed expression for the hard-scattering coefficient $C(z,M,\muf)$ in (\ref{factform}), which is given by the product of the solutions (\ref{CVsol}) and (\ref{sonice}) for the hard and soft functions. The result can be simplified by eliminating the anomalous dimension $\gamma^W$ using (\ref{gw}), and combining the Sudakov exponents using the relation
\begin{equation}
   S(\muh,\muf) - S(\mui,\muf) = S(\muh,\mui) 
   - a_\Gamma(\mui,\muf)\,\ln\frac{\muh}{\mui} \,.
\end{equation}
We find
\begin{eqnarray}\label{Cfinal}
   C(z,M,\muf) 
   &=& |C_V(-M^2,\muh)|^2\,U(M,\muh,\mui,\muf) \nonumber\\
   &&\times \frac{z^{-\eta}}{(1-z)^{1-2\eta}}\,\,
    \widetilde s_{\rm DY}\bigg( \ln\frac{M^2(1-z)^2}{\mui^2 z} 
     + \partial_\eta, \mui \bigg)\,
     \frac{e^{-2\gamma_E\eta}}{\Gamma(2\eta)} \,,
\end{eqnarray}
where $\eta=2a_\Gamma(\mui,\muf)$, and we have defined the evolution function
\begin{equation}\label{Udef}
   U(M,\muh,\mui,\muf) 
   = \left( \frac{M^2}{\muh^2} \right)^{-2a_\Gamma(\muh,\mui)}
   \exp\left[ 4S(\muh,\mui) - 2a_{\gamma^V}(\muh,\mui)
    + 4a_{\gamma^\phi}(\mui,\muf) \right] .
\end{equation}
As before, equation~(\ref{Cfinal}) is valid for $\eta>0$ ($\mui>\muf$). For negative $\eta$ ($\muf>\mui$), integrals of $\ln^n(1-z)/(1-z)^{1-2\eta}$ with test functions $f(z)$ must be defined using a subtraction at $z=1$ and analytic continuation in $\eta$.

We emphasize that the result (\ref{Cfinal}) is formally independent of the scales $\muh$ and $\mui$, at which the matching conditions for the hard and soft functions are evaluated. On the other hand, the hard-scattering kernel $C$ does depend on the factorization scale $\muf$, at which the PDFs are renormalized. In practice, a residual dependence on the matching scales arises when the perturbative expansions of the matching coefficients and anomalous dimensions are truncated, and this dependence can be used to estimate the remaining perturbative uncertainties. Setting the three scales $\muh$, $\mui$, and $\muf$ equal to each other in the resummed expression (\ref{Cfinal}), one can readily reproduce the leading singular terms for $z\to 1$ in the fixed-order perturbative QCD expression for the hard-scattering kernel. In this way we have obtained the two-loop corrections in (\ref{LLterms}).

\begin{table}
\centerline{\parbox{14cm}{\caption{\label{tab:counting}
Different approximation schemes for the evaluation of the resummed 
cross-section formulae}}}
\vspace{0.1cm}
\begin{center}
\begin{tabular}{ccrccc}
\hline
RG-impr.\ PT & Log.\ approx.\ & Accuracy $\sim\alpha_s^n L^k$
 & $\Gamma_{\rm cusp}$
 & $\gamma^V$, $\gamma^\phi$ & $C_V$, $\widetilde s_{\rm DY}$ \\
\hline\\[-0.4cm]
--- & LL & $k= 2n$ & 1-loop
 & tree-level & tree-level \\
LO & NLL & $2n-1\le k\le 2n$& 2-loop
 & 1-loop & tree-level \\
NLO & NNLL & $2n-3\le k\le 2n$ & 3-loop & 2-loop 
 & 1-loop \\
NNLO & NNNLL & $2n-5\le k\le 2n$ & 4-loop & 3-loop
 & 2-loop \\[0.2cm]
\hline
\end{tabular}
\end{center}
\end{table}

The final expression (\ref{Cfinal}) for the hard-scattering kernel can be evaluated at any desired order in resummed perturbation theory. 
Table~\ref{tab:counting} shows what is required to obtain different levels of accuracy. In this work we adopt the counting scheme of RG-improved perturbation theory, where at LO one includes all $O(1)$ terms, at NLO one includes all $O(\alpha_s)$ terms, etc. The large logarithm $\ln(\muh/\mui)$ is counted like $O(1/\alpha_s)$. In the literature on threshold resummation the alternative notation N$^{n+1}$LL is often used instead of N$^n$LO. The leading logarithmic (LL) approximation is listed only for completeness, as it misses some $O(1)$ terms.

In the following section we will perform a detailed numerical analysis of the Drell-Yan cross section and rapidity distribution. In most cases of phenomenological relevance the invariant mass of the Drell-Yan pair will be small compared with the center-of-mass energy, i.e.\ $\tau=M^2/s\ll 1$. Nevertheless, it is interesting to briefly consider the limit $\tau\to 1$, in which the need for threshold resummation is justified parametrically (see the discussion in Section~\ref{sec:pQCD}). In this case the convolution integrals in formula (\ref{dsigdM2}) for the Drell-Yan cross section can be performed analytically if a reasonably simple model for the PDFs near the endpoint is adopted. We parameterize the behavior near $x=1$ as
\begin{equation}\label{asymp}
   f_{q/N}(x,\muf) \big|_{x\to 1}
   = N_q(\muf)\,(1-x)^{b_q(\muf)} \Big[ 1 + O(1-x) \Big] \,,
\end{equation}
and similarly for the anti-quark distribution. It then follows that at leading power in $(1-y)$ the parton luminosity function defined in (\ref{ffdef}) is given by
\begin{equation}
   \ff(y,\muf) = 2 \sum_q e_q^2\,N_q(\muf)\,N_{\bar q}(\muf)\,
   (1-y)^{1+b_q+b_{\bar q}}\,
   \frac{\Gamma(1+b_q)\,\Gamma(1+b_{\bar q})}%
        {\Gamma(2+b_q+b_{\bar q})} \,.
\end{equation}
Furthermore, the sum over flavors is dominated by the terms with the smallest exponent $(b_q+b_{\bar q})$. For example, in $pp$ collisions at one finds $b_u+b_{\bar u}\approx 14.4$, $b_d+b_{\bar d}\approx 10.8$, and $b_s+b_{\bar s}\approx 16.0$ at $\mu_0=3$\,GeV \cite{Alekhin:2005gq}. These exponents increase by a flavor-independent amount $2a_\Gamma(\muf,\mu_0)$ when $\muf$ is raised to larger values, e.g., by about 0.4 at $\muf=8$\,GeV \cite{Becher:2006mr}. It follows that the leading behavior near the endpoint is due to the down-quark contribution.

Introducing the Drell-Yan $K$-factor as the ratio
\begin{equation}\label{Kdef}
   \frac{d\sigma}{dM^2}
   = K(M^2,\tau)\,\frac{d\sigma}{dM^2}\bigg|_{\rm LO} ,
\end{equation}
we then obtain at leading power
\begin{eqnarray}\label{KDY}
   K(M^2,\tau)
   &=& |C_V(-M^2,\muh)|^2\,U(M,\muh,\mui,\muf) \nonumber\\
   &&\times (1-\tau)^{2\eta}\,\,
    \widetilde s_{\rm DY}\Big( \ln\frac{M^2(1-\tau)^2}{\mui^2}
     +\partial_\eta,\mui \Big)\,
    \frac{e^{-2\gamma_E\eta}\,\Gamma(2+b_d+b_{\bar d})}
         {\Gamma(2+b_d+b_{\bar d}+2\eta)} \,. 
\end{eqnarray}
This may be compared with the $K$-factor for DIS at large Bjorken $x$ and momentum transfer 
\newpage\noindent
$Q^2=-q^2$, which has been derived in \cite{Becher:2006nr} and reads
\begin{eqnarray}
    K(Q^2,x)
   &=& |C_V(Q^2,\muh)|^2\,U(Q,\muh,\mui,\muf)\,
    \exp\left[ -2a_{\gamma^\phi}(\mui,\muf) \right] \nonumber\\
   &&\times (1-x)^\eta\,\,
    \widetilde j_{\rm DIS}\Big( \ln\frac{Q^2(1-x)}{\mui^2}
     +\partial_\eta,\mui \Big)\,
    \frac{e^{-\gamma_E\eta}\,\Gamma(1+b_u)}{\Gamma(1+b_u+\eta)} \,,
\end{eqnarray}
where in this case the dominant contribution comes from the valence up-quark, which has the smallest $b_q$ parameter, $b_u\approx 4.0$ at $\muf=3$\,GeV \cite{Alekhin:2005gq}. Obviously the structure of threshold logarithms is very similar in the two cases once we consider equal hard scales ($Q^2=M^2$) and compare the small parameter $(1-x)$ in DIS near the endpoint with the small parameter $(1-\tau)^2$ in Drell-Yan production. Differences arise from the following facts: (i) Drell-Yan production has time-like kinematics, whereas DIS probes the nucleon at space-like momentum transfer. This gives rise to a difference in the hard matching coefficients $|C_V|^2$ starting at one-loop order. (ii) The soft/jet functions in the two cases are different starting from one-loop order. (iii) The Drell-Yan cross section involves a convolution with two PDFs, whereas in DIS a single parton density appears. This explains the different coefficients in front of $a_{\gamma^\phi}$ and $\eta$ in the expressions for the $K$-factors. (iv) For the same reason, the resulting convolution integrals over the PDFs give rise to different expressions involving the $b_q$ and $b_{\bar q}$ exponents.

There is one more important piece of information that we can extract from the result (\ref{KDY}) for the Drell-Yan $K$-factor. As we have seen, the exponents $b_q$ and $b_{\bar q}$ take rather large values. Therefore, the arguments of the $\Gamma$-functions in (\ref{KDY}) contain the large quantity $(2+b_d+b_{\bar d})\approx 13$. It is straightforward to show that the derivative with respect to $\eta$ in the argument of the soft function has the effect of changing the argument of the logarithm as follows:
\begin{equation}
   \ln\frac{M^2(1-\tau)^2}{\mui^2} + \partial_\eta
   \to \ln\frac{M^2(1-\tau)^2}{\mui^2(2+b_d+b_{\bar d})^2} \,,
\end{equation}
up to $O(1)$ factors. It follows that a proper choice for the soft matching scale near $\tau\to 1$ is
\begin{equation}\label{magicmui}
   \mui\approx \frac{M(1-\tau)}{(2+b_d+b_{\bar d})} 
   \approx \frac{M(1-\tau)}{13} \,,
\end{equation}
which is an order of magnitude less than the naive choice $M(1-\tau)$. The fact that the fall-off of the parton densities strongly favors the large-$z$ region leads to a strong additional suppression of the effective soft scale.

\section{Phenomenological analysis}
\label{sec:numerics}

In this section we perform a detailed numerical analysis of our results. One of our goals is to study to what extent threshold resummation is important (or even justified) in processes where the invariant mass of the Drell-Yan pair is not very close to the center-of-mass energy. We have seen in Section~\ref{sec:pQCD} that for very small values of the ratio $\tau=M^2/s$ the threshold contributions are not parametrically enhanced. Even though empirically these terms still give rise to the dominant contributions to the cross section, there is no need to perform a resummation of the  threshold logarithms. On the other hand, the result (\ref{magicmui}) derived in the previous section shows that near the true endpoint the threshold logarithms are enhanced by two effects: the kinematic restriction of the $z$-integral to the small interval between $\tau$ and 1, as well as the dynamical enhancement from the strong fall-off of the parton densities. As a result, the appropriate value of the soft matching scale is an order of magnitude smaller than the naive choice $M(1-\tau)$. From our discussion so far it is not obvious how to interpolate between these two extreme cases. We now perform a detailed numerical study to assess the importance of resummation at intermediate values of $\tau$.

\subsection{Choices of the matching scales}

We begin with a discussion of the proper choice of the matching scales $\muh$ and $\mui$, using the convergence of the perturbative expansions of the matching coefficients $C_V$ and $\widetilde s_{\rm DY}$ in the resummed hard-scattering kernel (\ref{final}) as the primary guiding principle. While it is obvious that the hard matching scale should be chosen of order $M$, the choice of the soft scale is more problematic. Naively, based on the structure of the result (\ref{Cfinal}) one would expect that $\ln(\mui^2/M^2)$ should in some sense be identified with the ``average" value of $\ln[(1-z)^2/z]$. Unfortunately, however, the distribution in the variable $z$ is both singular at $z=1$ and not positive definite, so that it does not lend itself to a probabilistic interpretation. A simple way to avoid large logarithmic contributions would be to make the scale choice $\mui^2/M^2\propto[(1-z)^2/z]$ inside the $z$-integral. However, this would lead to Landau-pole singularities in the integrand and hence would upset the proper scale separation that is at the heart of our approach. 

\begin{figure}[t]
\begin{center}
\psfrag{x}[]{$\muh/M$}
\psfrag{y}[]{$c_n$}
\psfrag{w}[]{$c_1$}
\psfrag{z}[]{$c_2$}
\includegraphics[width=0.5\textwidth]{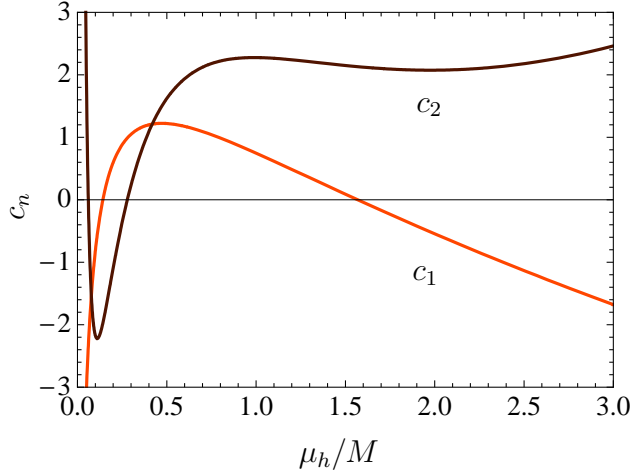}
\end{center}
\vspace{-0.5cm}
\caption{\label{fig:muhchoice}
Dependence of the first two expansion coefficients in the perturbative series for the hard function on the matching scale $\muh$.}
\end{figure}

Our approach to threshold resummation based on effective field-theory methods applied directly in momentum space provides a natural way of resolving this question. The matching scales $\muh$ and $\mui$ should be chosen such that the perturbative expansions of the Wilson coefficient functions $C_V$ and $\widetilde s_{\rm DY}$ in (\ref{Cfinal}) are well behaved. For the case of the soft function, this criterion should be applied {\em after\/} the integration over $z$ in (\ref{final}) has been performed. This is the essence of effective field theory: one performs matching calculations at scales where these calculations can be done in fixed-order perturbation theory, and use the renormalization group to perform the evolution (``running") between the different matching scales.

\begin{figure}[t]
\begin{center}
\psfrag{x}[]{\small $\mui$\,[GeV]}
\psfrag{y}[b]{\small $\alpha_s$ correction}
\includegraphics[width=0.45\textwidth]{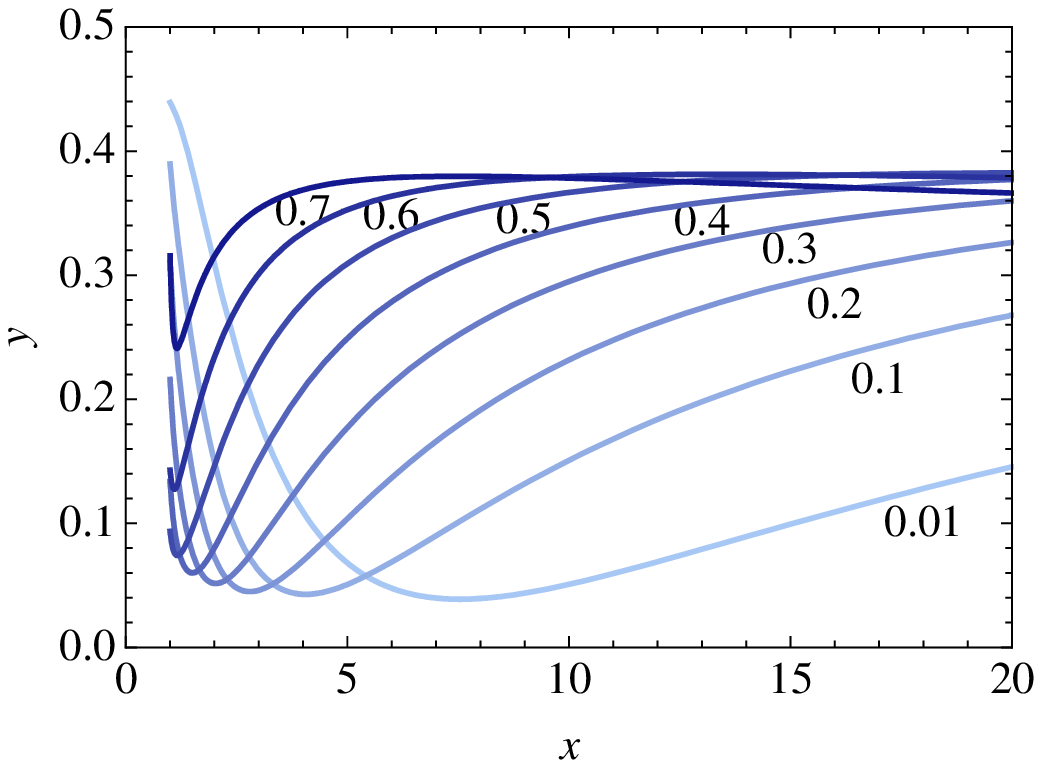}
\psfrag{y}[]{\small $\alpha_s^2$ correction}
\includegraphics[width=0.46\textwidth]{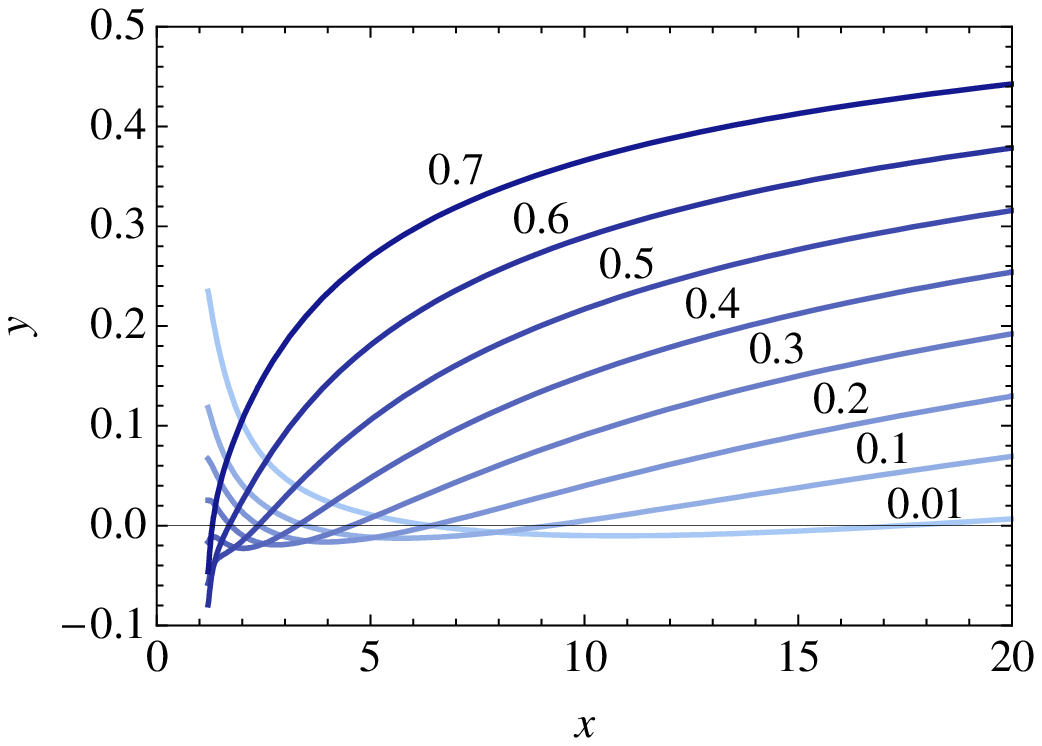}
\end{center}
\vspace{-0.5cm}
\caption{\label{fig:muichoice1}
Relative contributions to the Drell-Yan cross section $d\sigma/dM^2$ at $M=20$\,GeV arising from the one-loop (left) and two-loop (right) corrections to the soft function $\widetilde s_{\rm DY}$, as a function of the soft matching scale $\mui$. The curves are labeled by the corresponding values of $\tau=M^2/s$.}
\end{figure}

We begin by applying this criterion to the hard function
\begin{equation}\label{Hexp}
   H(M,\muh) = |C_V(-M^2,\muh)|^2
   = 1 + \sum_{n=1}^\infty c_n\Big( \frac{\muh}{M} \Big)
   \left[ \alpha_s(\muh) \right]^n .
\end{equation}
Figure~\ref{fig:muhchoice} shows the dependence of the expansion coefficients $c_1$ and $c_2$ on the ratio $\muh/M$. The one-loop coefficient $c_1$ vanishes for $\muh/M\approx 1.569$ and $\muh/M\approx 0.142$. The second solution is in a region of very small $\muh$, where the expansion coefficients vary strongly and where the $\ln^2(M^2/\muh^2)$ and $\ln(M^2/\muh^2)$ terms have opposite sign. We will thus discard it. In the region around the first solution the two-loop coefficient $c_2$ is stable and positive. In our numerical analysis we will vary $\muh$ between $M$ and $2M$, taking $\muh=3M/2$ as the default choice.

The matching scale $\mui$ must be determined separately for each process, since it is sensitive to the integration range of the $z$ variable (which depends on $\tau$ and $Y$) and to the shape of the PDFs. In Figure~\ref{fig:muichoice1}, we plot the relative contributions to the cross section $d\sigma/dM^2$ (normalized to the total cross section) arising from the one- and two-loop terms in the soft function $\widetilde s_{\rm DY}$ as a function of $\mui$. We choose $M=20$\,GeV and consider different values of $\tau=M^2/s$ between 0.01 and 0.7. The plots have been obtained by setting the factorization scale equal to $M$ and using MRST2004NNLO parton densities \cite{Martin:2004ir}. We have checked that virtually indistinguishable results are obtained when $\muf$ is varied by a factor of~2. Notice that with increasing $\tau$ values the regions where the one- and two-loop contributions are of modest size shift toward lower $\mui$ values. To be specific, we consider two criteria for a good convergence of the perturbative expansion (see the left plot in the figure):
\begin{itemize}
\item[I.] 
Starting from a high scale, we determine the value of $\mui$ at which the one-loop correction drops below 15\%.
\item[II.]
We choose the value of $\mui$ for which the one-loop contribution is minimal. 
\end{itemize}
Note that with either choice the two-loop corrections at the corresponding $\mui$ values are very small, indicating that the first two terms in the perturbation series for the soft function are well behaved for the same choice of scale. The same analysis can be repeated for different masses of the Drell-Yan pair. 

\begin{figure}
\begin{center}
\psfrag{x}[]{$\tau=M^2/s$}
\psfrag{y}[b]{$\mui/M$}
\includegraphics[width=0.5\textwidth]{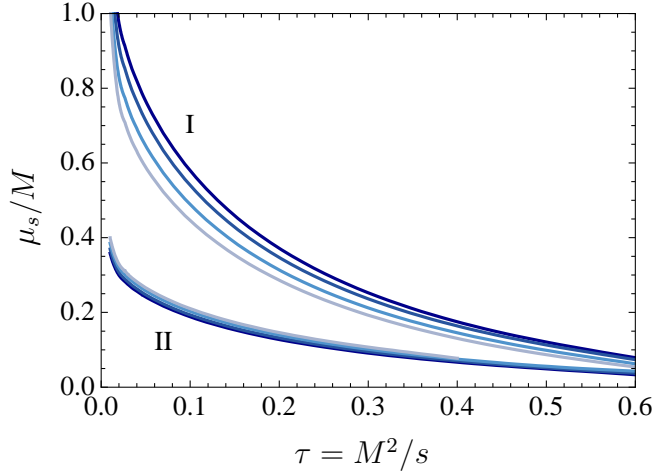}
\end{center}
\vspace{-0.5cm}
\caption{\label{fig:scaling}
Scale-setting results for the soft matching scale $\mui$ for different values of $\tau=M^2/s$ and Drell-Yan masses $M=100$, 50, 20, 10\,GeV. The lighter curves correspond to lower scales. The upper set of curves corresponds to convergence criterion I, the lower one to criterion II.}
\end{figure}

The resulting values for the soft scale $\mui$ determined using these convergence criteria are shown in Figure~\ref{fig:scaling}, where we consider the choices $M=10$, 20, 50, 100\,GeV. To a good approximation the curves for the ratio $\mui/M$ as a function of $\tau$ exhibit scaling, i.e., they are almost independent of $M$. We thus obtain a relation of the form $\mui/M=g(\tau)$. Small scaling violations arise from the scale dependence of the PDFs and of the running coupling in the perturbative expansion of the soft function. Our numerical results are well reproduced by the empirical functions
\begin{equation}\label{mui}
   \mui^{\rm I} = \frac{M(1-\tau)}{1+7\tau}
    \quad \mbox{and} \quad
   \mui^{\rm II} = \frac{M(1-\tau)}{\sqrt{6+150\tau}}
\end{equation}
for the two criteria. Their form should not be taken too seriously except to note that for $\tau\to 1$ both functions approach $\mui=\mbox{const.}\times M(1-\tau)$, as required by the resummation formula (\ref{KDY}) for the $K$-factor valid near the true endpoint. Indeed, the smallness of the constants in the two cases (0.125 and 0.080, respectively) is in good agreement with our estimate in (\ref{magicmui}). It results from the dynamical suppression provided by the PDFs. In the opposite limit $\tau\to 0$, both forms yield $\mui=\mbox{const.}\times M$ with an $O(1)$ constant, as required by the fact that there is only a single physical scale in this case. Below we will vary the soft scale between $\mui^{\rm I}$ and $\mui^{\rm II}$ and, somewhat arbitrarily, use the average of the two scales as the default choice.

The same analysis can be carried out for the double differential decay rate $d^2\sigma/dM dY$, and it leads to similar results. In general, one finds that at central rapidity ($Y=0$) the resulting values for the soft matching scale are higher by about 20\% than in the case of $d\sigma/dM^2$, while at larger rapidity they are lower, so that on average, after integration over $Y$, the results for the total cross section are reproduced.

\subsection{Scale dependence and impact of resummation}

We now proceed to study the stability of our resummed expression for the Drell-Yan cross section, using scale dependence as an estimator of yet unknown higher-order perturbative effects. We vary the matching scales $\muh$ and $\mui$ about the default values $\muh=1.5 M$ and $\mui=(\mui^{\rm I}+\mui^{\rm II})/2$ determined in the previous section. To simplify comparisons with the literature we adopt the conventional choice $\muf=M$ for the factorization scale, at which the PDFs are renormalized. From the point of view of effective field theory it would be more natural to choose a lower value for $\muf$, given that the cross section is sensitive to physics at scales much below the hard scale $M$. We will see, however, that our results are very stable with respect to variations of the factorization scale, so that the choice of the default value is not particularly important. Since our focus is on  the behavior of different perturbative approximations to the hard-scattering kernel, we use the same set of PDFs (MRST04NNLO) throughout the analysis.

\begin{figure}[t]
\begin{center}
\psfrag{la}[l]{\small ~$M<\muh<2M \phantom{\mu^{\rm I}}$}
\psfrag{x}[B]{\small $M^2/s$}
\psfrag{y}[]{\raisebox{0.5cm}{\small $K$}}
\begin{tabular}{cc}
\includegraphics[width=0.45\textwidth]{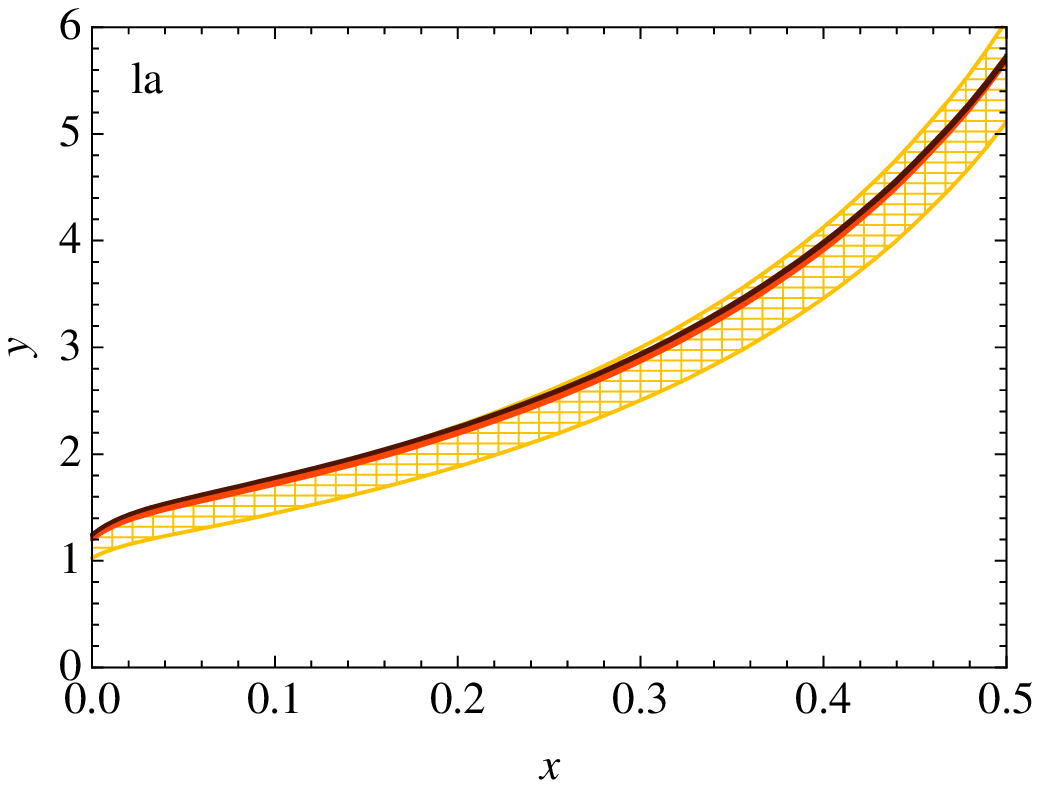} & 
\psfrag{la}[l]{\small ~$\mui^{\rm I}<\mui<\mui^{\rm II}$}
\includegraphics[width=0.45\textwidth]{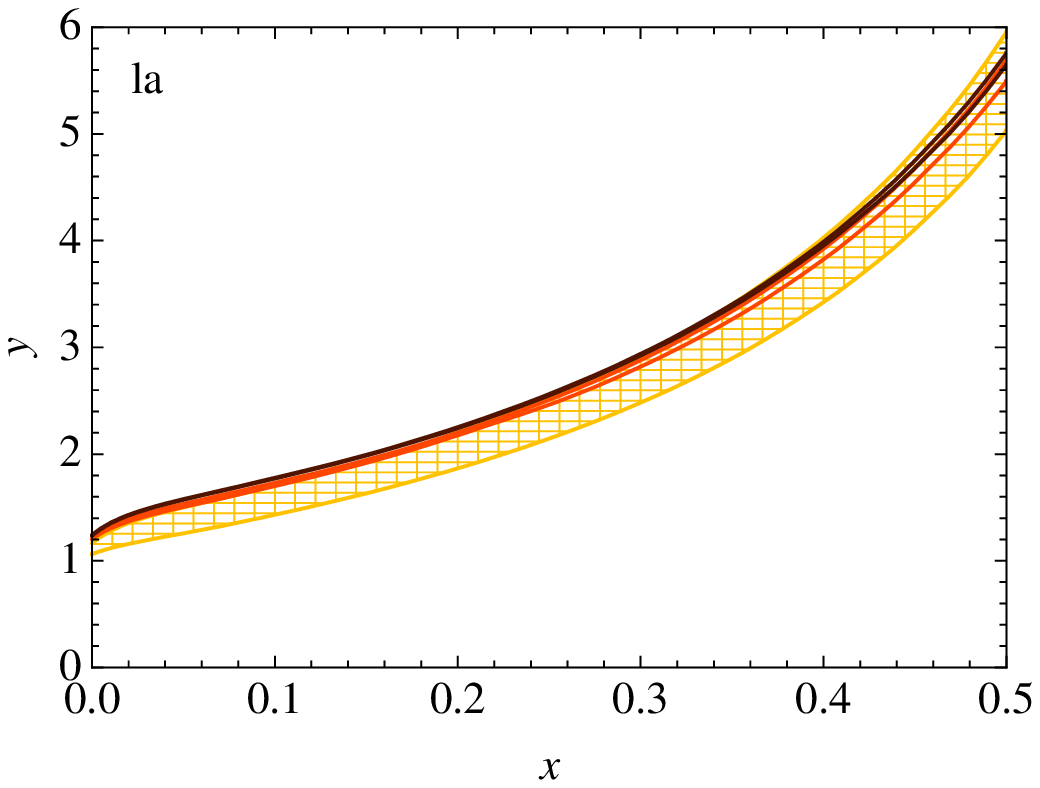}\\
\psfrag{la}[l]{\small ~$M/2<\muf<2M \phantom{\mu^{\rm I}}$}
\includegraphics[width=0.45\textwidth]{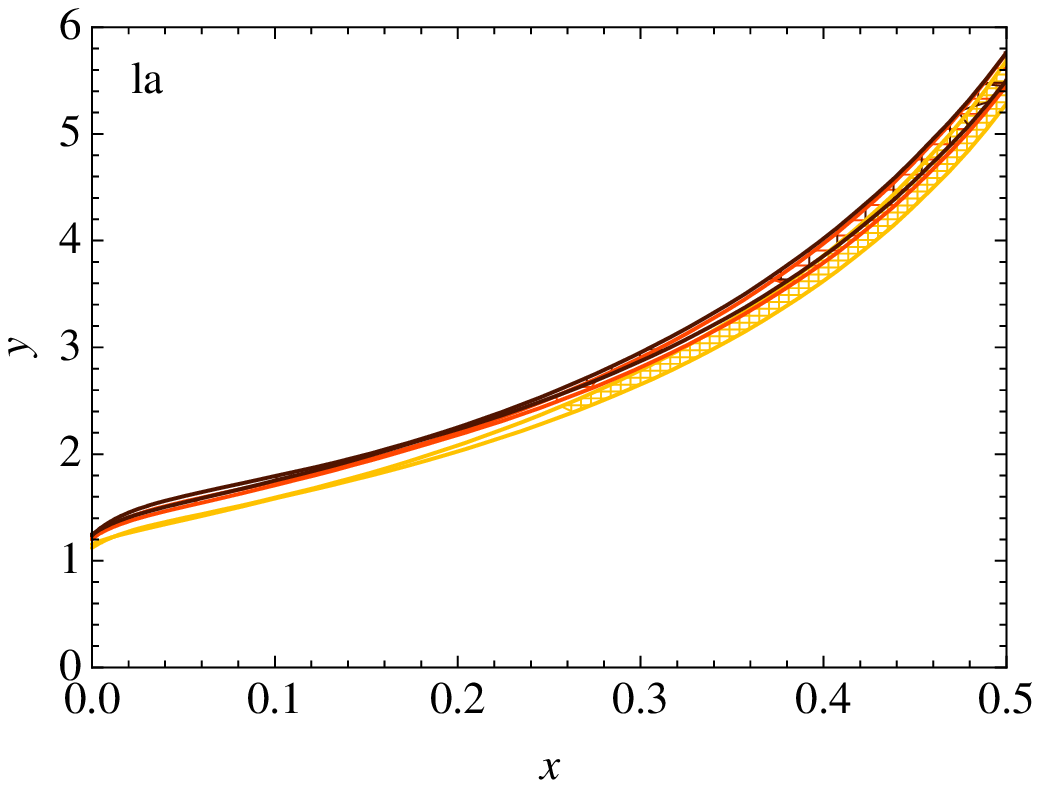} & 
\psfrag{la}[l]{\small ~$M/2<\muf<2M$ (fixed-order) 
 $\phantom{\mu^{\rm I}}$}
\includegraphics[width=0.45\textwidth]{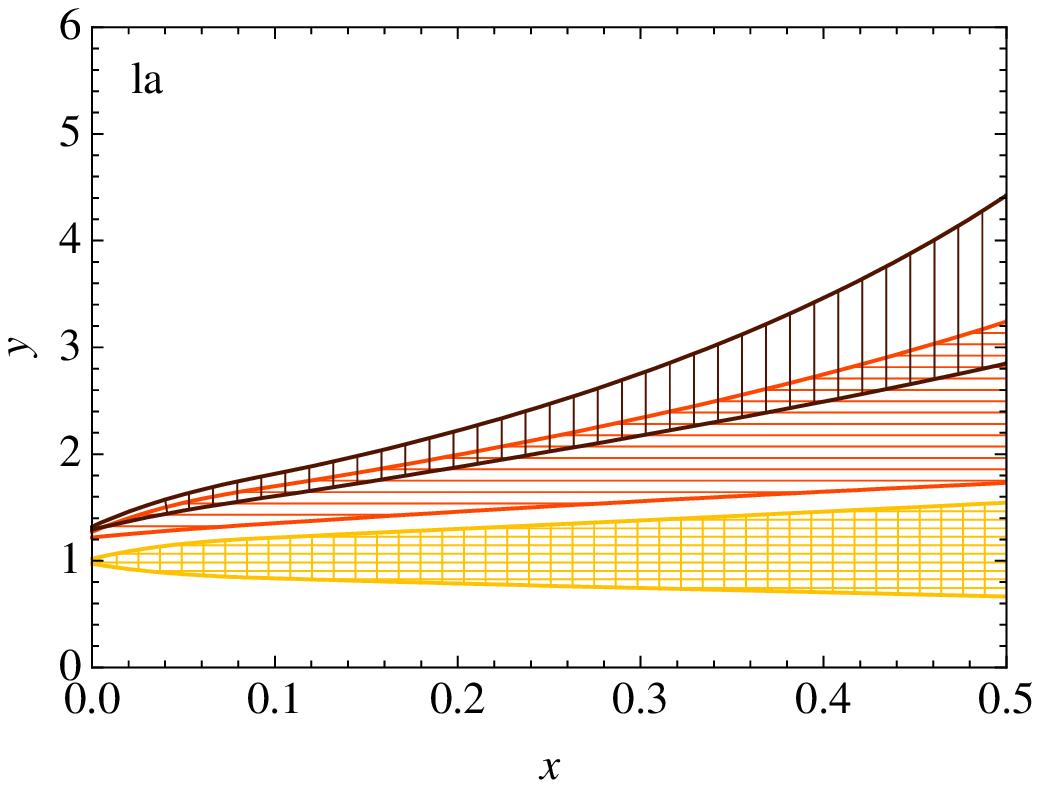} 
\end{tabular}
\end{center}
\vspace{-0.5cm}
\caption{\label{fig:scales}
Dependence of the resummed Drell-Yan cross section for $M=20$\,GeV on the scales $\muh$, $\mui$, and $\muf$. The bands show the $K$-factor obtained at LO (light), NLO (medium), and NNLO (dark). The last plot shows for comparison the $\muf$ dependence in fixed-order perturbation theory.}
\end{figure}

Our results are shown in the first three plots in Figure~\ref{fig:scales}, in which we study the Drell-Yan $K$-factor defined in (\ref{Kdef}) for $M=20$\,GeV and various values of $\tau=M^2/s$. In the calculation of the $K$-factor we keep the factorization scale $\muf=M$ fixed in the leading-order expression in the denominator, even when $\muf$ is varied in the numerator. For the time being we only include the leading terms in the $z\to 1$ limit, corresponding to the result (\ref{Cfinal}). Adding the small power-suppressed corrections would not change any of our conclusions. We observe an excellent convergence of the perturbative expansion for the cross section after resummation. The bands corresponding to the LO, NLO, and NNLO approximations overlap, and the dependence on the matching scales $\muh$ and $\mui$ becomes negligible beyond LO, indicating that the residual perturbative uncertainty is very small.

The third plot, showing the dependence on the factorization scale $\muf$, requires some comments. We first note that in the approximation where we resum the leading singular terms in the partonic cross section near $z\to 1$ but neglect power-suppressed terms, our results for the cross section are no longer strictly independent of the factorization scale $\muf$. In order to make them formally scale invariant one should add back the power-suppressed terms in fixed-order perturbation theory. Despite this, we find that the resummation greatly reduces the $\muf$ dependence compared with the fixed-order calculation, for which results are shown in the last plot in the figure. (The fixed-order results can be obtained by setting $\muh=\mui=\muf$ in the resummed expression.) The reason is that already the LO result after resummation compensates the leading scale dependence of the PDFs through the $\muf$ dependence of the functions $a_{\gamma^\phi}(\mui,\muf)$ and $\eta=2a_\Gamma(\mui,\muf)$ in (\ref{Cfinal}). Indeed, we see that for modest values $0<\tau<0.2$ the dependence on the factorization scale after resummation is almost absent already at LO. On the other hand, there is a sizable scale dependence in the fixed-order calculation even beyond LO. We emphasize that there is very little experimental information about the relevant parton densities (the sea-quark distributions, in particular) at large $x$ values, and as a consequence one cannot trust the numerical values of the $K$-factor at values $\tau>0.2$. Indeed, we find that the results for the cross section at $\tau=0.3$ obtained using different sets of PDFs (MRST04NNLO \cite{Martin:2004ir}, MRST01NNLO \cite{Martin:2002dr}, and CTEQ6.5 \cite{Tung:2006tb}) differ by a factor~4.

\begin{figure}[t]
\begin{center}
\begin{tabular}{cc}
\psfrag{la}[lt]{\small $\phantom{ab}\sqrt{s}=38.76$\,GeV}
\psfrag{M}[b]{\small $M$\,[GeV]}
\psfrag{K}[b]{\small $K$}
\includegraphics[width=0.455\textwidth]{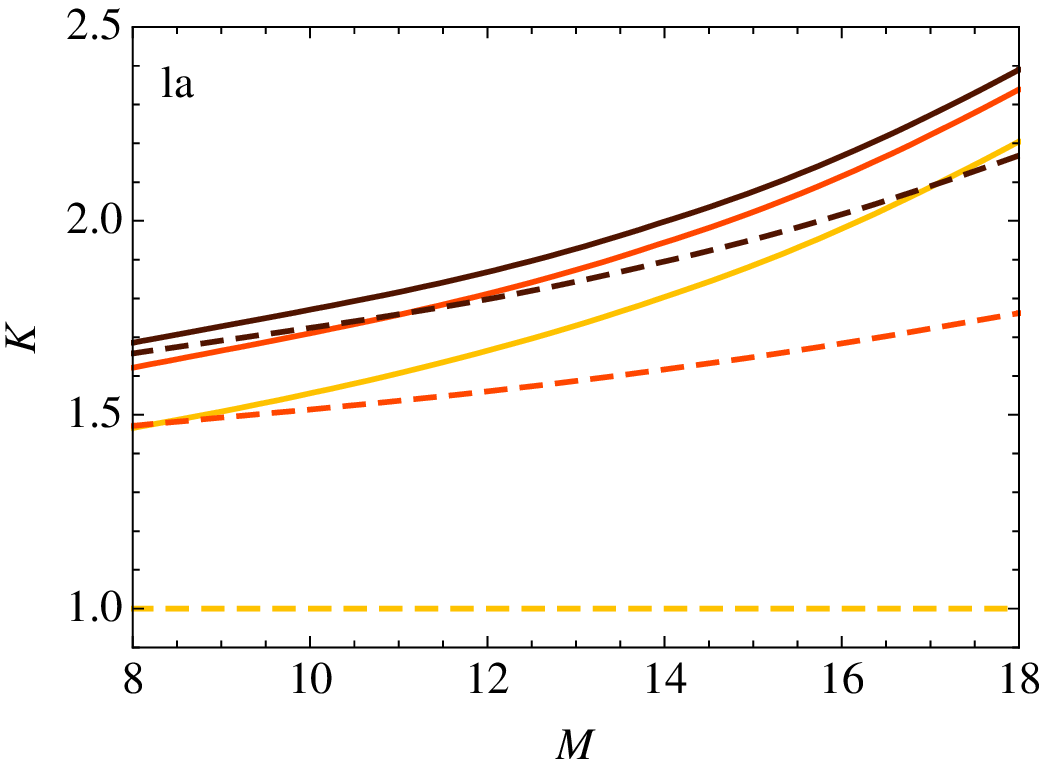} &
\psfrag{la}[lt]{\small $\phantom{ab}\sqrt{s}=14$\,TeV}
\psfrag{M}[B]{\small $M$\,[TeV]}
\psfrag{K}[B]{\small $K$}
\includegraphics[width=0.45\textwidth]{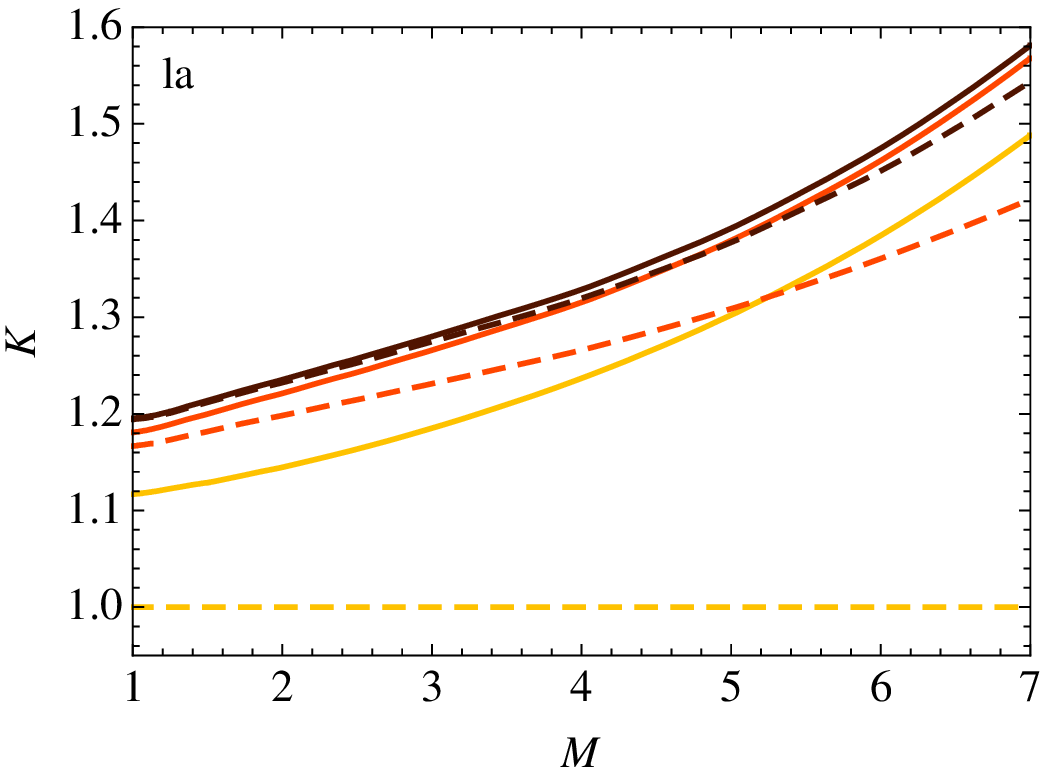} 
\end{tabular}
\end{center}
\vspace{-0.5cm}
\caption{\label{fig:plKfact}
Resummed (solid lines) versus fixed-order results (dashed lines) for the $K$-factor as a function of $M$. The light, medium, and dark lines correspond to LO, NLO, and NNLO, respectively. Default values are used for all scales.}
\end{figure}

In Figure~\ref{fig:fixedVSresummed}, we observed that the Drell-Yan
cross section is dominated by the singular threshold terms. We now assess whether the threshold contribution contains large logarithms which should be resummed. In Figure~\ref{fig:plKfact}, we compare the result for the total cross section obtained with the default values of the matching scales $\muh$ and $\mui$ to the evaluation of the threshold terms in fixed-order perturbation theory. The differences between the two sets of curves show the effect of the resummation. For illustration purposes we consider two examples. The first is the case of $pp$ collisions at $\sqrt{s}=38.76$\,GeV, corresponding to the energy of the fixed-target experiment E866/NuSea \cite{Webb:2003ps}. In this experiment Drell-Yan masses in the ranges 4.2--8.7\,GeV and 10.85--16.85\,GeV have been observed. As a second example we consider Drell-Yan production via a virtual photon at the LHC energy $\sqrt{s}=14$\,TeV (including the $Z^0$ channel would not alter our results for the $K$-factor significantly). The figure shows that resummation accelerates the convergence of the perturbative expansion. On the other hand, the plots also show that the most important logarithmic corrections are contained in the fixed-order NNLO results, at least for moderate lepton-pair masses. For large masses the resummation becomes more important. In the first example, the resummed cross section is about 8\% (28\%) larger than the fixed-order result at NNLO (NLO) for $M=16.85$\,GeV. At the LHC, on the other hand, resummation effects beyond the NNLO fixed-order result remain small even at very large values of $M$.

\begin{figure}
\begin{center}
\begin{tabular}{ccc}
\psfrag{x}[b]{\small $Y$}
\psfrag{y}[b]{\small $d^2\sigma/dM\,dY$~[pb/GeV]}
\psfrag{z}[l]{\small $\!\!\!\!M=8$\,GeV}
\includegraphics[height=0.33\textwidth]{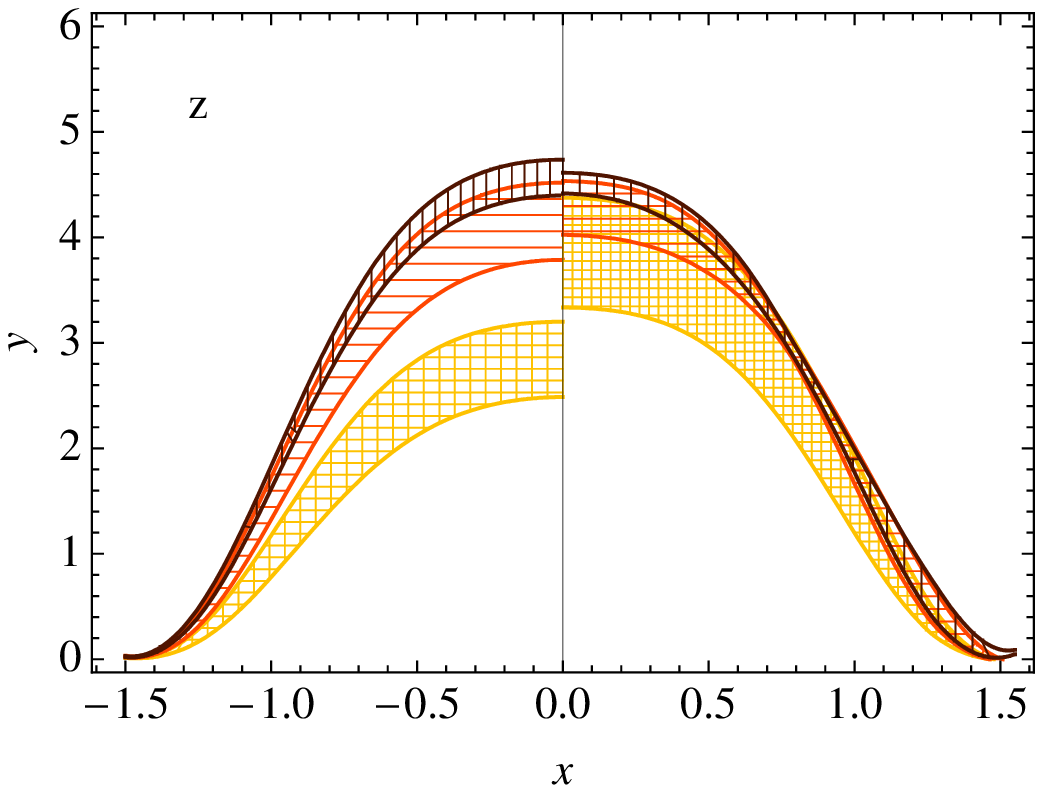} & &
\psfrag{x}[b]{\small $Y$}
\psfrag{y}[b]{\small $d^2\sigma/dM\,dY$~[pb/GeV]}
\psfrag{z}[l]{\small $\!\!\!\!M=16$\,GeV}
\includegraphics[height=0.33\textwidth]{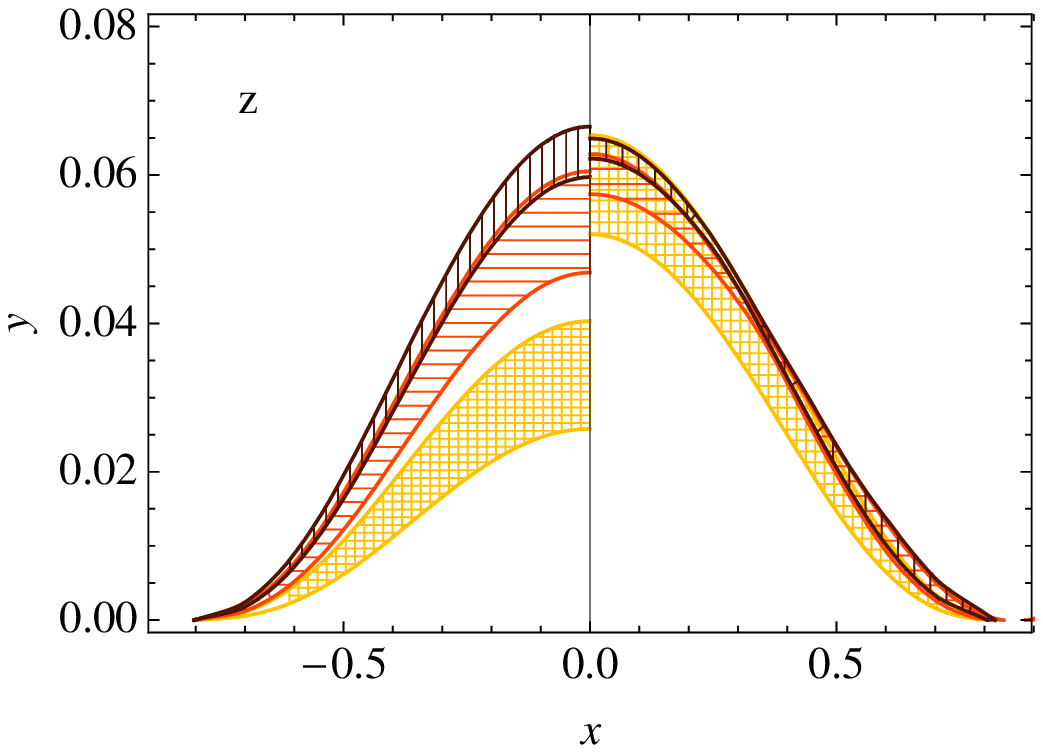} 
\end{tabular}
\end{center}
\vspace{-0.5cm}
\caption{\label{fig:rapidity}
Fixed-order ($Y<0$) versus resummed ($Y>0$) predictions for the rapidity distribution at $\sqrt{s}=38.76$\,GeV and two values of $M$, at different orders in perturbation theory. The bands reflect the combined scale dependence. LO bands are light, NLO bands are medium, NNLO bands are dark.}
\end{figure}

\subsection{Rapidity distribution and cross section at \boldmath $\sqrt{s}=38.76$\,GeV\unboldmath}

As a final application, we now return to the rapidity distribution in Drell-Yan production at $\sqrt{s}=38.76$\,GeV. As mentioned in the Introduction, in this case large resummation effects were found for $M=8$\,GeV \cite{Bolzoni:2006ky} even though $\tau\approx 0.04$ is very small. These effects were claimed to reduce the NLO fixed-order cross section by about 30\%. Fixed-order predictions for the rapidity distribution up to NNLO were discussed in \cite{Anastasiou:2003yy,Anastasiou:2003ds}. Here we present results for the two cases $M=8$ and 16\,GeV. In order to obtain the best possible predictions we 
combine our resummed result for the cross section with the power-suppressed terms calculated in fixed-order perturbation theory. In our approach this matching can be implemented in a straightforward way as follows:
\begin{equation}\label{match}
   \frac{d\sigma^{\rm combined}}{dM^2 dY} 
   = \left. \frac{d\sigma^{\rm thresh}}{dM^2 dY}
    \right|_{\muh,\mui,\muf}
   + \left( 
    \left. \frac{d\sigma^{\rm fixed\,order}}{dM^2 dY}\right|_{\muf} 
    - \left. \frac{d\sigma^{\rm thresh}}{dM^2 dY} 
    \right|_{\muh=\mui=\muf} \right) .
\end{equation}
In Figure~\ref{fig:rapidity}, we compare our RG-improved results with the fixed-order results, varying the scales over the ranges $M/2<\muf<2M$, $M<\muh<2M$, and $\mui^{\rm I}<\mui<\mui^{\rm II}$. The bands reflect the variations about the default value. In the fixed-order case only the first variation is relevant, while in the resummed case we add the individual variations in quadrature. 

\begin{table}[t]
\centerline{\parbox{14cm}{\caption{\label{tab:results}
Predictions for the Drell-Yan cross section $d\sigma/dM^2$ for $\sqrt{s}=38.76$\,GeV and two values of the invariant mass $M$ of the lepton pair. Units are pb/GeV$^2$ and fb/GeV$^2$, respectively.}}}
\vspace{0.1cm}
\begin{center}
\begin{tabular}{ccccc}
\hline
$M$ & Method & LO & NLO & NNLO \\
\hline\\[-0.4cm]
8\,GeV & combined & $0.436_{\,-0.071}^{\,+0.062}$
 & $0.493_{\,-0.014}^{\,+0.011}$ & $0.512_{\,-0.004}^{\,+0.002}$ \\
 & threshold & $0.436_{\,-0.071}^{\,+0.062}$ 
 & $0.482_{\,-0.015}^{\,+0.008}$ & $0.501_{\,-0.010}^{\,+0.009}$\\
 & fixed-order & $0.299_{\,-0.040}^{\,+0.051}$
 & $0.449_{\,-0.041}^{\,+0.051}$ & $0.505_{\,-0.025}^{\,+0.021}$
 \\[0.2cm] 
16\,GeV & combined & $1.49_{\,-0.21}^{\,+0.17}$
 & $1.61_{\,-0.04}^{\,+0.01}$ & $1.68_{\,-0.04}^{\,+0.01}$ \\
 & threshold & $1.49_{\,-0.21}^{\,+0.17}$
 & $1.59_{\,-0.04}^{\,+0.01}$ & $1.63_{\,-0.01}^{\,+0.01}$\\
 & fixed-order & $0.76_{\,-0.15}^{\,+0.21}$
 & $1.29_{\,-0.18}^{\,+0.19}$ & $1.57_{\,-0.12}^{\,+0.08}$ \\[0.2cm] 
\hline
\end{tabular}
\end{center}
\end{table}

We observe again that resummation significantly accelerates the convergence of the perturbative expansion. Moreover, even though in the resummed case we include the scale dependence from the variation of three different scales, the combined uncertainty at NLO and NNLO is significantly smaller than in the fixed-order case. Also, given the better overlap of the bands in the resummed case, our error estimates appear to be more conservative. As a final comment, we note that for $M=8$\,GeV the resummed results at NLO and NNLO are consistent within errors with the fixed-order results, indicating that threshold resummation is not an important effect. This is in stark contrast to the conclusion reached in \cite{Bolzoni:2006ky}. For the higher mass $M=16$\,GeV, the two NNLO bands are consistent with each other at central rapidity, but the resummed result is significantly higher than the fixed-order prediction for $Y\gtrsim 0.3$. For the integrated cross section at this value of $M$, threshold resummation enhances the fixed-order value by about 7\%. This can be seen from Table~\ref{tab:results}, which shows our final predictions for the integrated cross section $d\sigma/dM^2$. Besides the results obtained with and without resummation, we also give the contributions of the resummed threshold terms alone, corresponding to the first term in (\ref{match}).

\subsection{Resummation in moment space}

Traditionally, resummation is performed in moment rather than momentum space \cite{Sterman:1986aj,Catani:1989ne}. For the Drell-Yan cross section integrated over rapidity one takes moments in $\tau$ at fixed $M$:
\begin{equation}\label{momentdef}
   \sigma_N = \int_0^1\!d\tau\,\tau^{N-1}\,\frac{d\sigma}{dM^2} \,.
\end{equation}
For the moment-space analysis of the rapidity distribution one performs a Fourier transform in the rapidity in addition to taking moments in $\tau$ \cite{Sterman:2000pt,Bolzoni:2006ky}. In the following, we will restrict ourselves to the integrated cross section for simplicity. Using the representation (\ref{dsigdM2}), the cross section in moment space factorizes as
\begin{equation}\label{sigN}
   \sigma_N = \frac{4\pi\alpha^2}{3N_c M^4} 
   \sum_q e_q^2 \left[ f^{q/N_1}_{N+1}\,f^{\bar q/N_2}_{N+1}
    + (q\leftrightarrow\bar q) \right] C_{N+1}(M^2,\muf) \,,
\end{equation}
where the moments of the hard-scattering coefficient and the PDFs are defined in analogy with (\ref{momentdef}). In order to accomplish the resummation for the moments of the hard-scattering coefficient, we go back to expression (\ref{factform}) and use that \cite{Becher:2006qw}
\begin{equation}
   \int_0^1\!dz\,z^{N-1}\,S(M(1-z),\mu) 
   = \widetilde s_{\rm DY}\Big(\ln\frac{M^2}{\bar N^2\mu^2},\mu\Big)
   + O\Big(\frac{1}{N}\Big) \,,
\end{equation}
with $\bar N=e^{\gamma_E} N$. We then insert the solution to the RG evolution for the function $\widetilde s_{\rm DY}$, which as shown in \cite{Becher:2006mr} obeys a RG equation analogous to (\ref{gammaV}). This leads to
\begin{equation}\label{momentspaceCN}
   C_N(M^2,\muf)
   = |C_V(-M^2,\muh)|^2\,U(M,\muh,\mui,\muf)\,{\bar N}^{-2\eta}\,\,
   \widetilde s_{\rm DY}\Big(\ln\frac{M^2}{\bar N^2\mui^2},\mui\Big) 
   \,.
\end{equation}
Given the simple $N$ dependence of this result, we can transform it back to momentum space analytically using
\begin{eqnarray}\label{Mellininv}
   \frac{1}{2\pi i} \int_{c-i\infty}^{c+i\infty}\!dN\, 
   z^{-N} \bar N^{-2\eta} 
   &=& (-\ln z)^{-1+2\eta}\,\frac{e^{-2\gamma_E\eta}}{\Gamma(2\eta)}
    \nonumber\\
   &=& \sqrt{z}\,\frac{z^{-\eta}}{(1-z)^{1-2\eta}}\,
    \frac{e^{-2\gamma_E\eta}}{\Gamma(2\eta)}
    \left[ 1 + O\Big( (1-z)^2 \Big) \right]
\end{eqnarray}
and the fact that 
\begin{equation}
   {\bar N}^{-2\eta}\,\,
   \widetilde s_{\rm DY}\Big(\ln\frac{M^2}{\bar N^2\mui^2},\mui\Big) 
   = \widetilde s_{\rm DY}\Big(\ln\frac{M^2}{\mui^2}+\partial_\eta,
    \mui\Big)\,{\bar N}^{-2\eta} \,.
\end{equation}
Dropping the corrections of order $(1-z)^2$ in the last step in (\ref{Mellininv}), the inverted moment-space result can be written in a form that is identical to (\ref{Cfinal}) up to an overall factor $\sqrt{z}$, which amounts to a first-order power correction in the threshold region. In the limit $\muh=\mui=\muf$, in which the resummed expression reproduces the leading singular terms in the fixed-order result, the momentum-space formulation derived in the present work gives the singular distributions in the hard-scattering kernel in precisely the form in which they appear in (\ref{LLterms}) and (\ref{Piexpr}). The large-$N$ expansion in Mellin space, on the other hand, gives an expression that is obtained from this by the replacement
\begin{equation}
   \left[ \frac{L_z^n}{1-z} \right]_+ 
   \to \left[ \frac{\ln^n(M^2\ln^2 z/\muf^2)}{-\ln z} \right]_+ \,,
\end{equation}
with $L_z=\ln[M^2(1-z)^2/\muf^2 z]$. While the two expressions agree up to power-suppressed terms near the partonic threshold, they differ significantly for small $z$. This leads to large corrections when matching with the fixed-order results in cases where the small-$z$ region is relevant.

\begin{figure}[t]
\begin{center}
\psfrag{x}[B]{$N$}
\psfrag{y}[b]{$C_N/C_{N=1}^{\rm LO}$}
\includegraphics[width=0.5\textwidth]{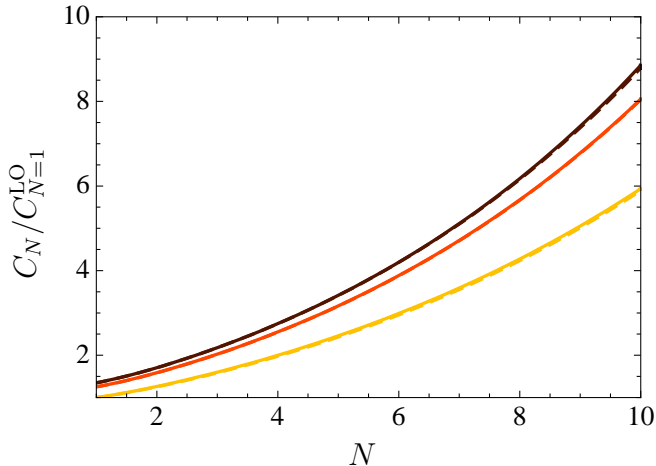} 
\end{center}
\vspace{-0.5cm}
\caption{\label{fig:comparisonMoment}
Comparison of the moment-space hard-scattering coefficient (\ref{momentspaceCN}) at $\muh=\muf=M$ and $\mui=M/N$ with $M=8$\,GeV (solid) with the expression (\ref{CNtrad}) used in the traditional approach \cite{Moch:2005ky} (dashed). The light, medium, and dark lines correspond to LO, NLO, and NNLO, respectively. The dashed curves are barely visible because they are almost on top of the solid lines.}
\end{figure}

In the traditional approach to resummation, the moment-space result is written as
\begin{equation}\label{CNtrad}
   C_N(M^2,\muf) = g_0(M^2,\muf)\,\exp[G_N(M^2,\muf)] 
   + O\Big(\frac{1}{N}\Big) \,,
\end{equation}
where $g_0$ collects the $N$-independent contributions. The resummation exponent has the form
\begin{equation}\label{GN}
   G_N(M^2,\muf) 
   = \int_0^1\!dz\,\frac{z^{N-1}-1}{1-z} \left[ 
   \int_{\muf^2}^{(1-z)^2 M^2}\!\frac{dk^2}{k^2}\,
   2A\left(\alpha_s(k)\right) + D\left(\alpha_s(M(1-z))\right)
   \right] .
\end{equation}
The coefficient $A(\alpha_s)$ is the cusp anomalous dimension, while $D(\alpha_s)$ is related to the anomalous dimension of the soft function. The resummation exponent evaluated to NNLO can be found in \cite{Moch:2005ky,Laenen:2005uz,Moch:2005ba,Ravindran:2005vv,Ravindran:2006cg}. Note that (\ref{GN}) is not well defined as it stands, since the coupling constant is integrated over the Landau pole. The resulting ambiguity corresponds to a spurious first-order power correction in $\Lambda_{\rm QCD}/M_X$, while on general grounds the cross sections is expected to receive power corrections starting at second order \cite{Beneke:1995pq}. For DIS, we have shown that for the scale choices $\muh=M$ and $\mui=M/N$, which are implicit in the traditional scheme, the two approaches to resummation are equivalent up to $1/N$ corrections, and we have derived the relationship between the anomalous dimensions in the effective theory and the resummation coefficients in the traditional approach \cite{Becher:2006mr}. With the same technique, we obtain
\begin{equation}
   e^{2\gamma_E\nabla}\,\Gamma(1+2\nabla)\,\frac{D(\alpha_s)}{2} 
   = \gamma_W(\alpha_s) + \nabla\ln\widetilde s_{\rm DY}(0,\mu)
   - \frac{e^{2\gamma_E\nabla}\,\Gamma(1+2\nabla)-1}{\nabla}\,
   \Gamma_{\rm cusp}(\alpha_s) \,,
\end{equation}
where $\alpha_s=\alpha_s(\mu)$, and $\nabla=d/d\ln\mu^2=[\beta(\alpha_s)/2]\,\partial/\partial\alpha_s$. Using this relation, we reproduce the perturbative expression for the function $D(\alpha_s)$ given in \cite{Moch:2005ky,Laenen:2005uz} up to third order in $\alpha_s^3$. In particular, for the first nonzero term we find
\begin{equation}
   D(\alpha_s) = \left( 2\gamma_1^W+2\pi^2 C_F\beta_0 \right)
   \left( \frac{\alpha_s}{4\pi} \right)^2 + O(\alpha_s^3) \,.
\end{equation}

In Figure~\ref{fig:comparisonMoment}, we compare our moment-space expression (\ref{momentspaceCN}) for the hard-scattering kernel $C_N$ evaluated with the scale choices $\muh=M$ and $\mui=M/N$ with expression (\ref{CNtrad}) evaluated as described in \cite{Moch:2005ky}. The results are almost indistinguishable, demonstrating that the $1/N$-suppressed differences between the two formulations are numerically very small, and that the traditional approach to threshold resummation can be viewed as a special case of our RG-based framework.

\begin{figure}[t]
\begin{center}
\begin{tabular}{cc}
\psfrag{la}[tl]{\small $M=20$\,GeV}
\psfrag{x}[B]{\small$M^2/s$}
\psfrag{y}[b]{\small $K$}
\includegraphics[width=0.45\textwidth]{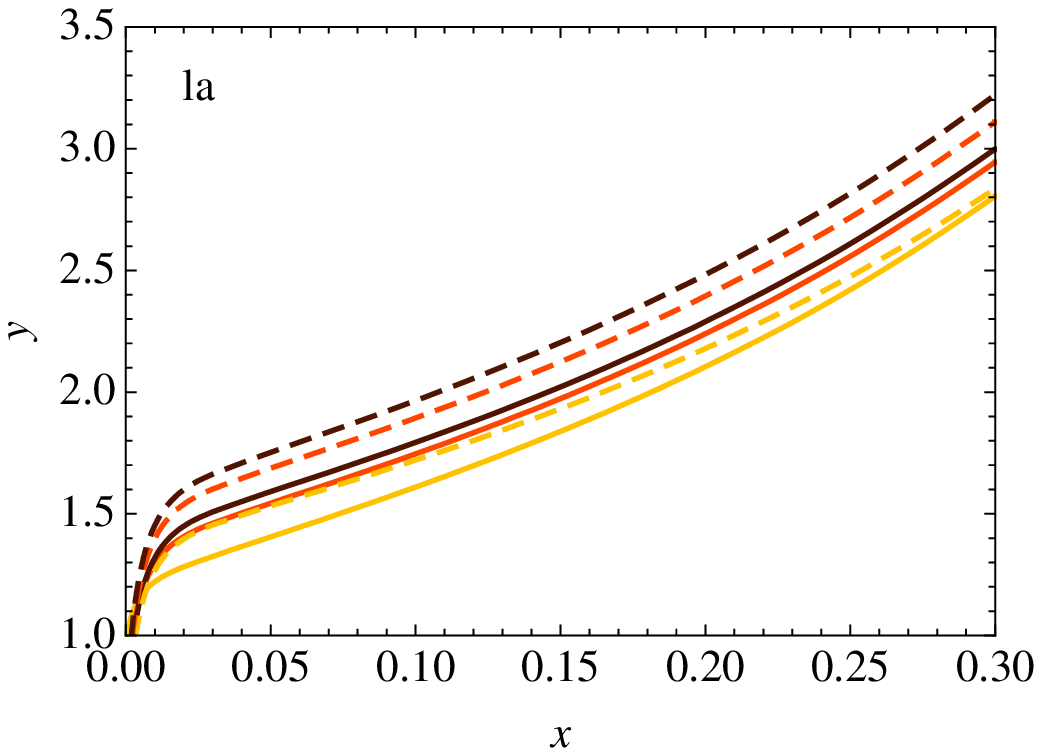} &
\psfrag{la}[tl]{\small $M=20$\,GeV}
\psfrag{x}[B]{\small $M^2/s$}
\psfrag{y}[b]{\small $K$}
\includegraphics[width=0.45\textwidth]{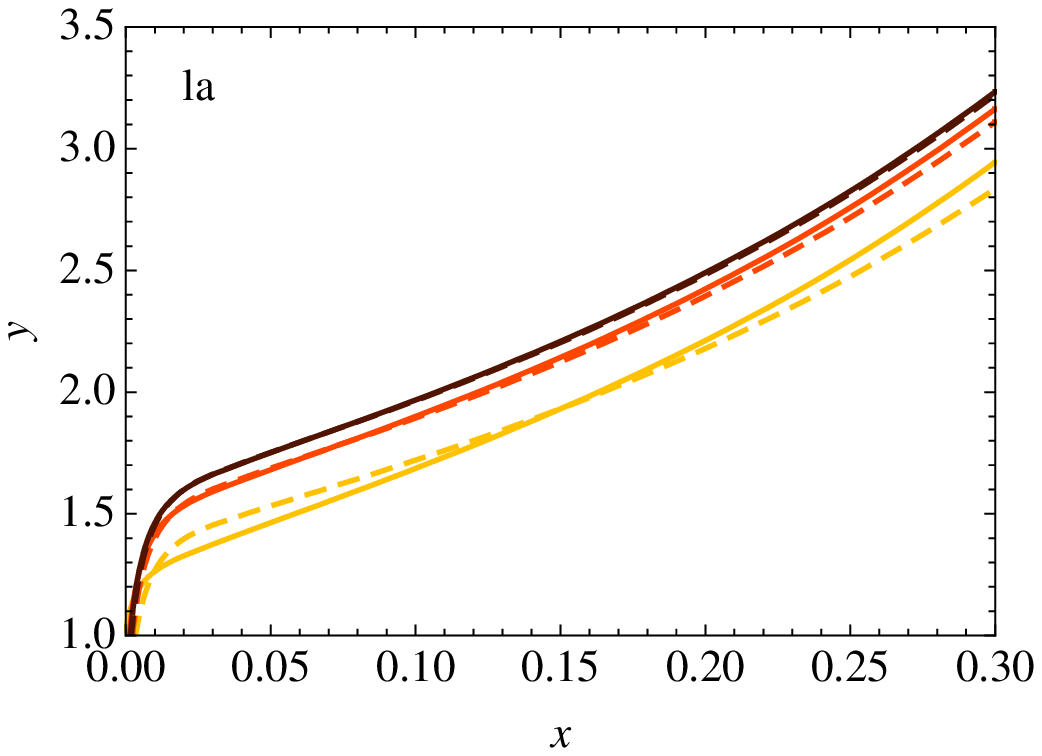} 
\end{tabular}
\end{center}
\vspace{-0.5cm}
\caption{\label{fig:compMoment}
Left: Results for the Drell-Yan $K$-factor obtained using threshold resummation in momentum-space (solid) and moment space (dashed). The default scale choices $\mu_h=1.5 M$, $\mui=(\mui^{\rm I}+\mui^{\rm II})/2$, and $\mu_f=M$ are used in both cases.
Right: Moment-space results for the $K$-factor obtained from (\ref{momentspaceCN}) with scale choices $\muh=1.5M$ and $\mui=M/N$ (dashed) and the default choices $\muh=1.5M$ and $\mui=(\mui^{\rm I}+\mui^{\rm II})/2$ (solid). The light, medium, and dark lines correspond to LO, NLO, and NNLO, respectively.}
\end{figure}

The fact that the moment-space resummation approach differs from the momentum-space approach by first-order corrections in $(1-z)$ leads to visible numerical effects. In order to illustrate this we take the inverse Mellin transformation of (\ref{sigN}) to compute the Drell-Yan $K$-factor. For simplicity, we fit the polynomial form $\ff(y)=y^a(1-y)^b(1+cy+dy^2)$ to the parton luminosity function defined in (\ref{ffdef}), so that the product of the moments of the PDFs in (\ref{sigN}) can be evaluated in closed form. The first plot in Figure~\ref{fig:compMoment} shows that for the same choices of matching scales the moment-space approach leads to somewhat larger results for the $K$-factor than the momentum-space approach, and that the convergence is slightly better in the second case.

In the traditional moment-space approach based on expression (\ref{GN}) one implicitly makes the scale choice $\mui=M/N$, which leads to a Landau pole at $N\sim M/\Lambda_{\rm QCD}$ in the Mellin inversion. To see how the results obtained with this choice compare with our default results we adopt the so-called minimal prescription \cite{Catani:1996yz}, which amounts to choosing an inversion contour in the $N$-plane that does not include this pole. When this is done, we find that the moment-space formula (\ref{momentspaceCN}) evaluated with $\muh=1.5M$ and $\mui=M/N$ gives numerical results similar to those found using our default scale choices once we go beyond leading order. This is illustrated in the right plot in Figure~\ref{fig:compMoment}.

\section{Conclusions}

We have presented a detailed analysis of threshold resummation for Drell-Yan production. Instead of the conventional Mellin moment formalism, we have performed the resummation directly in momentum space, using RG evolution. We have obtained analytic expressions for the resummed hard-scattering kernels that are free of the Landau-pole ambiguities inherent in the traditional approach. After deriving the necessary matching coefficients and anomalous dimensions from known perturbative results in the literature, we have performed the resummation for the cross section and rapidity distribution at NNLO in RG-improved perturbation theory, corresponding to N$^3$LL accuracy.

The main goal of this work was to address the question to what extent resummation can be phenomenologically relevant given that in all practical applications the true threshold region is experimentally not accessible, because the parton distributions fall off very steeply at large $x$. It has been argued in the literature that precisely this fall-off forces the Drell-Yan process to the partonic threshold region, in which large logarithmic corrections arise. Our analysis confirms the existence of this effect and quantifies its importance. In the true endpoint region, we find that the scale of soft radiation is an order of magnitude smaller than the naive estimate. The extra suppression factor is given by the sum of the exponents of the fall-off of the quark and anti-quark distributions near $x=1$. On the other hand, for very small lepton-pair masses this effect becomes inoperative because the fall-off of the parton distributions is not very steep at low $x$. An analysis of the convergence of perturbation theory in the intermediate region shows that resummation effects become relevant for $M$ larger than about $0.4\sqrt{s}$. At the largest measured Drell-Yan masses resummation effects can thus be significant. In the case of the experiment E866/NuSea, a fixed-target experiment with $pp$ collisions at $\sqrt{s}=38.76$\,GeV, these effects increase the NNLO (NLO) cross section by 7\% (25\%) at $M=16$\,GeV.

We do not confirm the recent claim of large negative resummation effects for the rapidity distribution measured in the E866/NuSea experiment at $M=8$\,GeV \cite{Bolzoni:2006ky}. We find that the fixed-order NNLO threshold contribution is positive and a good approximation to the full NNLO correction. Higher-order terms beyond NNLO turn out to be negligible, so that it is unnecessary to resum them. To check whether the large effect seen in \cite{Bolzoni:2006ky} could be an artifact of the moment-space formalism, we have applied our approach also in moment space and showed that it encompasses the traditional moment-space approach. We find that the two methods give compatible results, and that in both schemes resummation increases the cross section. The Drell-Yan cross section is dominated by the contribution of the threshold region even at small masses of the lepton pair. For example, for $M=8$\,GeV the threshold terms give rise to $93\%$ of the NLO correction to the total cross section and $97\%$ of the NNLO contribution. If this is a more generic feature of hard cross sections, then threshold resummation techniques provide an efficient way to obtain the dominant part of the higher-order corrections. 

Perhaps the most important outcome of our analysis is a quantitative understanding of the emergence of an effective physical scale characterizing the soft radiation in the Drell-Yan process. This scale is generated through an intricate interplay of dynamical and kinematical effects. It will be interesting to explore how the same mechanism affects other collider processes such as Higgs production at the Tevatron and LHC. Furthermore, it would be useful to analyze  resummation for $t\bar t$-production, which is phenomenologically relevant, since the Tevatron now produces top-quark pairs with quite high invariant masses. There was some controversy on how to best perform the resummmation in this case \cite{Catani:1996dj,Berger:1996ad,Catani:1996yz,Berger:1997gz}. It would be interesting to revisit the issue using effective field-theory methods.

\subsection*{Acknowledgments}

We are grateful to Lance Dixon, Keith Ellis, Frank Petriello, and Werner Vogelsang for useful discussions. M.N.\ likes to thank the Fermilab Theory Group for hospitality and support as a Frontier Fellow. The research of T.B.\ was supported by the Department of Energy under Grant DE-AC02-76CH03000. The research of G.X.\ was supported by the National Science Foundation under Grant PHY-0355005. Fermilab is operated by the Fermi Research Alliance under contract with the U.S.\ Department of Energy.

\newpage
\begin{appendix}

\section{Two-loop coefficients of the leading singular terms}
\label{app:a}

The leading singular terms in the fixed-order perturbative expressions for the hard-scattering kernels $C_{ij}$ can be derived by using the perturbative expansions of the hard and soft functions in (\ref{factform}). In this way, we obtain for the coefficients $P_i(z)$ in (\ref{LLterms})
\begin{eqnarray}\label{Piexpr}
   P_F(z) &=& \delta(1-z) \left[ \frac98\,L^2
    + \left( - \frac{93}{16} + \frac{3\pi^2}{4} + 5\zeta_3 \right) L 
    + \frac{511}{64} - \frac{33\pi^2}{16} + \frac{23\pi^4}{120} 
    - \frac{15\zeta_3}{4} \right] \nonumber\\
   &&\mbox{}+ 16\zeta_3 \left[ \frac{1}{1-z} \right]_+
    + \left( - L^2 + 3L - 8 - \frac{2\pi^2}{3} \right)
    \left[ \frac{L_z}{1-z} \right]_+
    + \left[ \frac{L_z^3}{1-z} \right]_+ \,, \nonumber\\
   P_A(z) &=& \delta(1-z) \left[ - \frac{11}{16}\,L^2 
    + \left( \frac{193}{48} - \frac{11\pi^2}{36} 
    - \frac{3\zeta_3}{2} \right) L
    - \frac{1535}{192} + \frac{47\pi^2}{36} - \frac{23\pi^4}{720} 
    + \frac{43\zeta_3}{12} \right] \nonumber\\
   &&\mbox{}+ \left( - \frac{101}{27} + \frac{11\pi^2}{18} 
    + \frac{7\zeta_3}{2} \right) \left[ \frac{1}{1-z} \right]_+
    + \left( \frac{67}{18} - \frac{\pi^2}{6} \right)
    \left[ \frac{L_z}{1-z} \right]_+
    - \frac{11}{12} \left[ \frac{L_z^2}{1-z} \right]_+ \,, \nonumber\\
   P_f(z) &=& \delta(1-z) \left[ \frac14\,L^2
    + \left( - \frac{17}{12} + \frac{\pi^2}{9} \right) L 
    + \frac{127}{48} - \frac{4\pi^2}{9} + \frac{\zeta_3}{3} \right]
    \nonumber\\
   &&\mbox{}+ \left( \frac{28}{27} - \frac{2\pi^2}{9} \right)
    \left[ \frac{1}{1-z} \right]_+
    - \frac{10}{9} \left[ \frac{L_z}{1-z} \right]_+
    + \frac13 \left[ \frac{L_z^2}{1-z} \right]_+ \,.
\end{eqnarray}
As before, we use the abbreviations $L=\ln(M^2/\muf^2)$ and $L_z=\ln[M^2(1-z)^2/\muf^2 z]$. If desired, the plus distributions involving powers of $L_z$ can be reduced to distributions of the form $[\ln^n(1-z)/(1-z)]_+$ using the identity
\begin{equation}
   \left[ f(z)\,g(z) \right]_+
   = f(z) \left[ g(z) \right]_+ - \delta(1-z) \int_0^1\!dz'\,f(z')
   \left[ g(z') \right]_+ \,.
\end{equation}

\section{Matching coefficients and anomalous dimensions}
\label{Appendix:RGfunctions}

For completeness we list the perturbative expansions of the various matching coefficients and anomalous dimensions required to evaluate our RG-improved result (\ref{Cfinal}) at NNLO. 

\subsection{Two-loop matching coefficients}

The matching conditions for the Wilson coefficient $C_V$ evaluated at time-like momentum transfer can be obtained by analytic continuation from the corresponding expression valid at space-like momentum transfer. Using the known two-loop result for the on-shell QCD form factor \cite{Matsuura:1987wt,Matsuura:1988sm,Gehrmann:2005pd,Moch:2005id}, we find \cite{Becher:2006mr}
\begin{equation}
   C_V(-M^2-i\epsilon,\mu) = 1 + \frac{C_F\alpha_s}{4\pi}
    \left( - L^2 + 3L - 8 + \frac{\pi^2}{6} \right)
    + C_F \left( \frac{\alpha_s}{4\pi} \right)^2 \left[
    C_F H_F + C_A H_A + T_F n_f H_f \right] ,
\end{equation}
where
\begin{eqnarray}
   H_F &=& \frac{L^4}{2} - 3L^3
    + \left( \frac{25}{2} - \frac{\pi^2}{6} \right) L^2
    + \left( - \frac{45}{2} - \frac{3\pi^2}{2} + 24\zeta_3 \right) L
    + \frac{255}{8} + \frac{7\pi^2}{2} - \frac{83\pi^4}{360} 
    - 30\zeta_3 \,, \nonumber\\
   H_A &=& \frac{11}{9}\,L^3
    + \left( - \frac{233}{18} + \frac{\pi^2}{3} \right) L^2
    + \left( \frac{2545}{54} + \frac{11\pi^2}{9} - 26\zeta_3 \right) L 
    \nonumber\\
   &&\mbox{}- \frac{51157}{648} - \frac{337\pi^2}{108} + \frac{11\pi^4}{45}
    + \frac{313}{9}\,\zeta_3 \,, \nonumber\\
   H_f &=& - \frac49\,L^3 + \frac{38}{9}\,L^2 
    + \left( - \frac{418}{27} - \frac{4\pi^2}{9} \right) L
    + \frac{4085}{162} + \frac{23\pi^2}{27} + \frac49\,\zeta_3 \,,
\end{eqnarray}
and $L=\ln(M^2/\mu^2)-i\pi$. This result agrees with the corresponding expression given in \cite{Idilbi:2006dg}. 

The matching condition for the function $\widetilde s_{\rm DY}$ can be derived most easily using relation (\ref{wonderful}), which relates it to the perturbative expansion of the position-space Wilson loop $\hat W_{\rm DY}$ at time-like separation. We can then use the explicit two-loop expression for the position-space Wilson loop obtained in \cite{Belitsky:1998tc}. To this end, we have obtained the fourth-order coefficient in the expansion of a certain Appell hypergeometric function,
\begin{equation}
   F_2\Big( \begin{array}{c} 1,1+\epsilon,-2\epsilon \\[-0.15cm]
            2+\epsilon,1-2\epsilon \end{array}\,\Big|\,1,1 \Big)
   = - \frac{1+\epsilon}{2\epsilon} \left( 
    1 - \frac{\pi^2}{3}\,\epsilon^2 - 14\zeta_3\epsilon^3 
    - \frac{5\pi^4}{18}\,\epsilon^4 + \dots \right) .
\end{equation}
The resulting two-loop expression for the soft function reads
\begin{equation}
   \widetilde s_{\rm DY}(L,\mu)
   = 1 + \frac{C_F\alpha_s}{4\pi} 
   \left( 2L^2 + \frac{\pi^2}{3} \right)
   + C_F \left( \frac{\alpha_s}{4\pi} \right)^2
   \left[ C_F W_F + C_A W_A + T_F n_f W_f \right] ,
\end{equation}
where
\begin{eqnarray}
   W_F &=& 2L^4 + \frac{2\pi^2}{3}\,L^2 + \frac{\pi^4}{18} 
    = \frac12 \left( 2L^2 + \frac{\pi^2}{3} \right)^2 , \nonumber\\
   W_A &=& - \frac{22}{9}\,L^3
    + \left( \frac{134}{9} - \frac{2\pi^2}{3} \right) L^2
    + \left( - \frac{808}{27} + 28\zeta_3 \right) L
    + \frac{2428}{81} + \frac{67\pi^2}{54} - \frac{\pi^4}{3}
    - \frac{22}{9}\,\zeta_3 \,, \nonumber\\
   W_f &=& \frac89\,L^3 - \frac{40}{9}\,L^2 + \frac{224}{27}\,L
    - \frac{656}{81} - \frac{10\pi^2}{27} + \frac89\,\zeta_3 \,.
\end{eqnarray}
Note that the $C_F^n$ terms exponentiate, which is a consequence of the non-abelian exponentiation theorem for Wilson loops \cite{Gatheral:1983cz,Frenkel:1984pz}. Our result for the function $\widetilde s_{\rm DY}$ agrees with a corresponding expression entering in the moment-space resummation approach studied in \cite{Idilbi:2006dg}. 

\subsection{Three-loop anomalous dimensions}

Here we list expressions for the anomalous dimensions and the QCD 
$\beta$-function, quoting all results in the $\overline{{\rm MS}}$ 
renormalization scheme. We define the expansion coefficients of the anomalous dimensions and the QCD $\beta$-function as
\begin{eqnarray}
   \Gamma_{\rm cusp}(\alpha_s) &=& \Gamma_0\,\frac{\alpha_s}{4\pi}
    + \Gamma_1 \left( \frac{\alpha_s}{4\pi} \right)^2
    + \Gamma_2 \left( \frac{\alpha_s}{4\pi} \right)^3 + \dots \,,
    \nonumber\\
   \beta(\alpha_s) &=& -2\alpha_s \left[
    \beta_0\,\frac{\alpha_s}{4\pi}
    + \beta_1 \left( \frac{\alpha_s}{4\pi} \right)^2
    + \beta_2 \left( \frac{\alpha_s}{4\pi} \right)^3 + \dots \right] ,
\end{eqnarray}
and similarly for the other anomalous dimensions. 

The expansion of the cusp anomalous dimension $\Gamma_{\rm cusp}$ to two-loop order was obtained some time ago \cite{Korchemskaya:1992je}, while recently the three-loop coefficient has been calculated in \cite{Moch:2004pa}. For the four-loop coefficient $\Gamma_3$ we use the Pad\'e approximant $\Gamma_3=\Gamma_2^2/\Gamma_1$. The results are
\begin{eqnarray}
   \Gamma_0 &=& 4 C_F \,, \nonumber\\
   \Gamma_1 &=& 4 C_F \left[ \left( \frac{67}{9} 
    - \frac{\pi^2}{3} \right) C_A - \frac{20}{9}\,T_F n_f \right] ,
    \nonumber\\
   \Gamma_2 &=& 4 C_F \Bigg[ C_A^2 \left( \frac{245}{6} 
    - \frac{134\pi^2}{27}
    + \frac{11\pi^4}{45} + \frac{22}{3}\,\zeta_3 \right) 
    + C_A T_F n_f  \left( - \frac{418}{27} + \frac{40\pi^2}{27}
    - \frac{56}{3}\,\zeta_3 \right) \nonumber\\
   &&\mbox{}+ C_F T_F n_f \left( - \frac{55}{3} + 16\zeta_3 \right) 
    - \frac{16}{27}\,T_F^2 n_f^2 \Bigg] \,, \nonumber\\
   \Gamma_3 &\approx& 7849,~ 4313, ~ 1553 \quad \mbox{for} \quad
    n_f = 3,\,4,\,5 \,.
\end{eqnarray}
The anomalous dimension $\gamma^V$ can be determined up to three-loop order from the partial three-loop expression for the on-shell quark form factor in QCD, which has recently been obtained in \cite{Moch:2005id}. We find
\begin{eqnarray}
   \gamma_0^V &=& -6 C_F \,, \nonumber\\
   \gamma_1^V &=& C_F^2 \left( -3 + 4\pi^2 - 48\zeta_3 \right)
    + C_F C_A \left( - \frac{961}{27} - \frac{11\pi^2}{3} 
    + 52\zeta_3 \right)
    + C_F T_F n_f \left( \frac{260}{27} + \frac{4\pi^2}{3} \right) ,
    \nonumber\\
   \gamma_2^V &=& C_F^3 \left( -29 - 6\pi^2 - \frac{16\pi^4}{5}
    - 136\zeta_3 + \frac{32\pi^2}{3}\,\zeta_3 + 480\zeta_5 \right) 
    \nonumber\\
   &&\mbox{}+ C_F^2 C_A \left( - \frac{151}{2} + \frac{410\pi^2}{9}
    + \frac{494\pi^4}{135} - \frac{1688}{3}\,\zeta_3
    - \frac{16\pi^2}{3}\,\zeta_3 - 240\zeta_5 \right) \nonumber\\
   &&\mbox{}+ C_F C_A^2 \left( - \frac{139345}{1458} - \frac{7163\pi^2}{243}
    - \frac{83\pi^4}{45} + \frac{7052}{9}\,\zeta_3
    - \frac{88\pi^2}{9}\,\zeta_3 - 272\zeta_5 \right) \nonumber\\
   &&\mbox{}+ C_F^2 T_F n_f \left( \frac{5906}{27} - \frac{52\pi^2}{9} 
    - \frac{56\pi^4}{27} + \frac{1024}{9}\,\zeta_3 \right) 
    \nonumber\\
   &&\mbox{}+ C_F C_A T_F n_f \left( - \frac{34636}{729}
    + \frac{5188\pi^2}{243} + \frac{44\pi^4}{45} 
    - \frac{3856}{27}\,\zeta_3 \right) \nonumber\\
   &&\mbox{}+ C_F T_F^2 n_f^2 \left( \frac{19336}{729} 
    - \frac{80\pi^2}{27} - \frac{64}{27}\,\zeta_3 \right) .
\end{eqnarray}
The anomalous dimension $\gamma^\phi$ is know to three-loop order from the NNLO calculation of the Altarelli-Parisi splitting functions \cite{Moch:2004pa}. The expansion coefficients are
\begin{eqnarray}
   \gamma_0^\phi &=& 3 C_F \,, \nonumber\\
   \gamma_1^\phi 
   &=& C_F^2 \left( \frac{3}{2} - 2\pi^2 + 24\zeta_3 \right) 
    + C_F C_A \left( \frac{17}{6} + \frac{22\pi^2}{9} 
    - 12\zeta_3 \right)
    - C_F T_F n_f \left( \frac{2}{3} + \frac{8\pi^2}{9} \right) ,
    \nonumber\\
   \gamma_2^\phi
   &=& C_F^3 \left( \frac{29}{2} + 3\pi^2 + \frac{8\pi^4}{5} 
    + 68\zeta_3 - \frac{16\pi^2}{3}\,\zeta_3 - 240\zeta_5 \right)
    \nonumber\\
   &&\mbox{}+ C_F^2 C_A \left( \frac{151}{4} - \frac{205\pi^2}{9}
    - \frac{247\pi^4}{135} + \frac{844}{3}\,\zeta_3
    + \frac{8\pi^2}{3}\,\zeta_3 + 120\zeta_5 \right) \nonumber\\
   &&\mbox{}+ C_F^2 T_F n_f \left( - 46 + \frac{20\pi^2}{9}
    + \frac{116\pi^4}{135} - \frac{272}{3}\,\zeta_3 \right) 
    \nonumber\\
   &&\mbox{}+ C_F C_A^2 \left( - \frac{1657}{36}
    + \frac{2248\pi^2}{81} - \frac{\pi^4}{18} 
    - \frac{1552}{9}\,\zeta_3 + 40\zeta_5 \right) \nonumber\\
   &&\mbox{}+ C_F C_A T_F n_f \left( 40 - \frac{1336\pi^2}{81}
    + \frac{2\pi^4}{45} + \frac{400}{9}\,\zeta_3 \right) \nonumber\\
   &&\mbox{}+ C_F T_F^2 n_f^2 \left( - \frac{68}{9} 
    + \frac{160\pi^2}{81} - \frac{64}{9}\,\zeta_3 \right) .
\end{eqnarray}
Using these results, one can compute the expansion coefficients for the anomalous dimension $\gamma^W$ of the Drell-Yan soft function from the relation $\gamma^W=2\gamma^\phi+\gamma^V$. This yields
\begin{eqnarray}
   \gamma_0^W &=& 0 \,, \nonumber\\
   \gamma_1^W &=& C_F C_A \left( - \frac{808}{27} + \frac{11\pi^2}{9} 
    + 28\zeta_3 \right)
    + C_F T_F n_f \left( \frac{224}{27} - \frac{4\pi^2}{9} \right) .
\end{eqnarray}
We do not list the three-loop coefficient.

Finally, the expansion coefficients for the QCD $\beta$-function to four-loop order are
\begin{eqnarray}
   \beta_0 &=& \frac{11}{3}\,C_A - \frac43\,T_F n_f \,, \nonumber\\
   \beta_1 &=& \frac{34}{3}\,C_A^2 - \frac{20}{3}\,C_A T_F n_f
    - 4 C_F T_F n_f \,, \\
   \beta_2 &=& \frac{2857}{54}\,C_A^3 + \left( 2 C_F^2
    - \frac{205}{9}\,C_F C_A - \frac{1415}{27}\,C_A^2 \right) T_F n_f
    + \left( \frac{44}{9}\,C_F + \frac{158}{27}\,C_A 
    \right) T_F^2 n_f^2 \,, \nonumber\\
   \beta_3 &=& \frac{149753}{6} + 3564\zeta_3
    - \left( \frac{1078361}{162} + \frac{6508}{27}\,\zeta_3 
    \right) n_f
    + \left( \frac{50065}{162} + \frac{6472}{81}\,\zeta_3 
    \right) n_f^2
    + \frac{1093}{729}\,n_f^3 \,. \nonumber
\end{eqnarray}
The value of $\beta_3$ is taken from \cite{vanRitbergen:1997va} and corresponds to $N_c=3$ and $T_F=\frac12$. 

\subsection{Renormalization-group functions}

We now give the perturbative expansions of the functions $S$ and $a_\Gamma$ defined in (\ref{RGEsols}), working consistently at NNLO in RG-improved perturbation theory. At this order we need to keep terms through $O(\alpha_s^2)$ in the final expressions. The resulting expression for $a_\Gamma$ is given by
\begin{eqnarray}\label{asol}
   a_\Gamma(\nu,\mu)
   &=& \frac{\Gamma_0}{2\beta_0}\,\Bigg\{
    \ln\frac{\alpha_s(\mu)}{\alpha_s(\nu)}
    + \left( \frac{\Gamma_1}{\Gamma_0} - \frac{\beta_1}{\beta_0} 
    \right) \frac{\alpha_s(\mu) - \alpha_s(\nu)}{4\pi} \nonumber\\ 
   &&\mbox{}+ \left[ \frac{\Gamma_2}{\Gamma_0}
    - \frac{\beta_2}{\beta_0} - \frac{\beta_1}{\beta_0}
    \left( \frac{\Gamma_1}{\Gamma_0} - \frac{\beta_1}{\beta_0} 
    \right) \right]
    \frac{\alpha_s^2(\mu) - \alpha_s^2(\nu)}{32\pi^2} + \dots \Bigg\}
    \,.
\end{eqnarray}
Similar expressions with the $\Gamma_i$ replaced by the coefficients $\gamma_i^V$ or $\gamma_i^\phi$ hold for the functions $a_{\gamma^V}$ and $a_{\gamma^\phi}$, respectively. The NNLO expression for the Sudakov exponent $S$ is more complicated. It reads \cite{Becher:2006mr}
\begin{eqnarray}
   S(\nu,\mu) 
   &=& \frac{\Gamma_0}{4\beta_0^2}\,\Bigg\{
    \frac{4\pi}{\alpha_s(\nu)} \left( 1 - \frac{1}{r} - \ln r \right)
    + \left( \frac{\Gamma_1}{\Gamma_0} - \frac{\beta_1}{\beta_0}
    \right) (1-r+\ln r) + \frac{\beta_1}{2\beta_0} \ln^2 r \nonumber\\
   &&\mbox{}+ \frac{\alpha_s(\nu)}{4\pi} \Bigg[ 
    \left( \frac{\beta_1\Gamma_1}{\beta_0\Gamma_0} 
    - \frac{\beta_2}{\beta_0} \right) (1-r+r\ln r)
    + \left( \frac{\beta_1^2}{\beta_0^2} 
    - \frac{\beta_2}{\beta_0} \right) (1-r)\ln r \nonumber\\
   &&\hspace{1.0cm}
    \mbox{}- \left( \frac{\beta_1^2}{\beta_0^2} 
    - \frac{\beta_2}{\beta_0}
    - \frac{\beta_1\Gamma_1}{\beta_0\Gamma_0} 
    + \frac{\Gamma_2}{\Gamma_0} \right) \frac{(1-r)^2}{2} \Bigg]
    \nonumber\\
   &&\mbox{}+ \left( \frac{\alpha_s(\nu)}{4\pi} \right)^2 \Bigg[
    \left( \frac{\beta_1\beta_2}{\beta_0^2} 
    - \frac{\beta_1^3}{2\beta_0^3}
    - \frac{\beta_3}{2\beta_0} + \frac{\beta_1}{\beta_0}
    \left( \frac{\Gamma_2}{\Gamma_0} - \frac{\beta_2}{\beta_0}
    + \frac{\beta_1^2}{\beta_0^2} 
    - \frac{\beta_1\Gamma_1}{\beta_0\Gamma_0} \right) 
    \frac{r^2}{2} \right) \ln r \nonumber\\
   &&\hspace{1.0cm}
    \mbox{}+ \left( \frac{\Gamma_3}{\Gamma_0} 
    - \frac{\beta_3}{\beta_0}
    + \frac{2\beta_1\beta_2}{\beta_0^2} 
    + \frac{\beta_1^2}{\beta_0^2}
    \left( \frac{\Gamma_1}{\Gamma_0} - \frac{\beta_1}{\beta_0} \right)
    - \frac{\beta_2\Gamma_1}{\beta_0\Gamma_0}
    - \frac{\beta_1\Gamma_2}{\beta_0\Gamma_0} \right) 
    \frac{(1-r)^3}{3} \nonumber\\
   &&\hspace{1.0cm}
    \mbox{}+ \left( \frac{3\beta_3}{4\beta_0} 
    - \frac{\Gamma_3}{2\Gamma_0} + \frac{\beta_1^3}{\beta_0^3}
    - \frac{3\beta_1^2\Gamma_1}{4\beta_0^2\Gamma_0}
    + \frac{\beta_2\Gamma_1}{\beta_0\Gamma_0}
    + \frac{\beta_1\Gamma_2}{4\beta_0\Gamma_0}
    - \frac{7\beta_1\beta_2}{4\beta_0^2} \right) (1-r)^2 \nonumber\\
   &&\hspace{1.0cm}
    \mbox{}+ \left( \frac{\beta_1\beta_2}{\beta_0^2} 
    - \frac{\beta_3}{\beta_0}
    - \frac{\beta_1^2\Gamma_1}{\beta_0^2\Gamma_0}
    + \frac{\beta_1\Gamma_2}{\beta_0\Gamma_0} \right) \frac{1-r}{2}
    \Bigg] + \dots \Bigg\} \,,
\end{eqnarray}
where $r=\alpha_s(\mu)/\alpha_s(\nu)$. Whereas the three-loop anomalous dimensions and $\beta$-function are required in (\ref{asol}), the expression for $S$ also involves the four-loop coefficients $\Gamma_3$ and $\beta_3$.

\newpage
\section{Cut diagrams in the Keldysh formalism}
\label{sec:keldysh}

\begin{figure}
\begin{center}
\psfrag{p}[t]{$P$}
\psfrag{q}[t]{$P'$}
\includegraphics[width=0.45\textwidth]{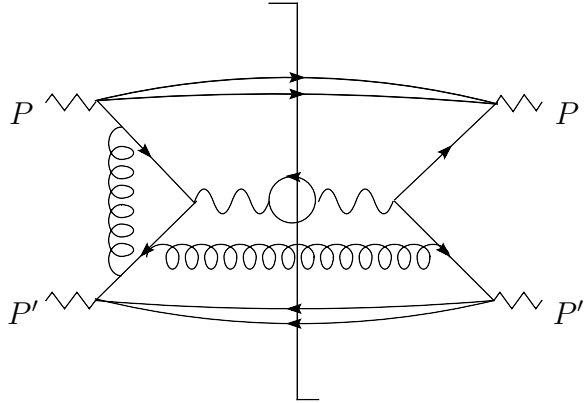}
\end{center}
\vspace{-0.5cm}
\caption{Example of a cut diagram contribution to the Drell-Yan process.
\label{fig:cutDiagram}}
\end{figure}

To perform the effective theory analysis of the process, we would like to have a path-integral definition of the hadronic quantity of interest. In the case of DIS (and also for inclusive $B$-decays), one considers the discontinuity of forward matrix elements of time-ordered products of the electroweak currents and studies their factorization properties. The Drell-Yan cross section cannot be written in a similar form. The reason is that not all cuts of the relevant Feynman graphs correspond to the same physical process. For example, in addition to the contribution from the cut indicated in Figure~\ref{fig:cutDiagram}, the discontinuity of the same diagram also gets a contribution from a cut through the triangle loop on the left, which describes $p+\bar p\to X$ with a virtual lepton pair.

\begin{figure}
\begin{center}
\psfrag{t}[t]{$t$}
\includegraphics[width=0.4\textwidth]{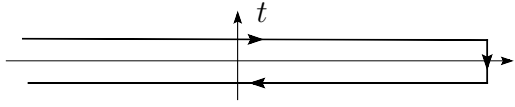}
\end{center}
\vspace{-0.5cm}
\caption{Contour in the complex $t$-plane along which the Lagrangian is integrated to obtain the action in the Keldysh formalism. (To dampen the oscillatory behavior of the path integrand, the entire contour should further be rotated clockwise by a small angle.)
\label{fig:contour}}
\end{figure}

A path-integral method for the direct evaluation of cut diagrams is the Keldysh (or time-loop) formalism \cite{Schwinger:1960qe,Keldysh:1964ud}. Let us illustrate the method for a scalar field theory with field $\phi(x)$. Instead of the usual action, where one integrates over time from $t=-\infty\dots\infty$, one considers a path integral with an action
\begin{equation}
   S(\phi) = \int_{\cal C}dt\,{\cal L}(\phi) \,,
\end{equation}
where the contour $\cal C$ first runs from $t=-\infty+i\delta \dots \infty+i\delta$ and then back, with a negative imaginary part, along the contour shown in Figure~\ref{fig:contour}. When evaluating expectation values of fields, one needs to specify on which part of the contour the fields reside. We denote the fields living on the first half of the contour by $\phi_+(x)\equiv\phi(x+i\delta)$ and the ones on the second half by $\phi_-(x)\equiv\phi(x-i\delta)$. The path-integral expectation values of fields correspond to path-ordered vacuum expectation values. Since the fields $\phi_+$ are positioned at earlier points along the contour, they are to the right of all fields $\phi_-$. Furthermore path-ordering translates into anti-time-ordering on the second part of the contour, so that we find
\begin{eqnarray}\label{expectation}
   &&\int{\cal D}\phi\,\phi_+(x_1)\dots\phi_+(x_n)\, 
    \phi_-(x_1)\dots\phi_-(x_m) \exp\left\{ i S(\phi)\right\}
    \nonumber\\
   &=& \langle 0|\,\overline{\rm\bf T}
    \left\{ \phi(x_1)\dots\phi(x_m) \right\} 
    {\rm\bf T} \left\{ \phi(x_1)\dots\phi(x_n) \right\} |0\rangle \,.
\end{eqnarray}
Here $\overline{\rm\bf T}$ denotes anti-time-ordering. Note that under a field redefinition $\phi(x)\to f(\phi(x))\,\phi(x)$, the plus and minus fields transform separately $\phi_\pm(x)\to f(\phi_\pm(x))\,\phi_\pm(x)$. This is relevant when one uses a field redefinition to decouple the soft gluon fields from the quark fields in the current operator.

The Keldysh formalism is useful, because it gives a path integral formulation of squared amplitudes, e.g.
\begin{eqnarray}
   &&\sum_X (2\pi)^4 \delta(P_X-P)\,
    |\langle X|\,{\rm\bf T}
    \left\{ \phi(x_1)\dots\phi(x_n) \right\} |0\rangle |^2 \nonumber\\
   &=& \int\!d^4x\,e^{-iP\cdot x}\,\langle 0|\,\overline{\rm\bf T} 
    \left\{ \phi(x_1+x)\dots\phi(x_n+x) \right\} 
    {\rm\bf T} \left\{ \phi(x_1)\dots\phi(x_n) \right\} 
    |0\rangle \,,
\end{eqnarray}
which is then rewritten as an expectation value of $\phi_\pm$ fields using (\ref{expectation}).

For our perturbative analysis, we need the Feynman rules to calculate matrix elements of the form (\ref{expectation}). The rules are simple: those for the field $\phi_+$ are the usual Feynman rules and those for $\phi_-$ are the complex conjugate of the usual rules. Since all interactions are local, there are no vertices involving both $\phi_+$ and $\phi_-$ fields. The only connection is the cut propagator
\begin{equation}
   \langle 0|\,\phi_-(x)\,\phi_+(0)\,|0\rangle 
   = \int\frac{d^4p}{(2\pi)^4}\,e^{-ip\cdot x}\,(2\pi)\, 
   \delta(p^2-m^2)\,\theta(p^0) \,. 
\end{equation}
This expression is familiar from the Cutkosky rules \cite{Cutkosky:1960sp} used to extract the contribution of a given cut to an ordinary Feynman diagram.

Considering the case of the Drell-Yan cross section in (\ref{textbook}), the relevant quantity to consider in the Keldysh formalism is the current product $J^\mu_-(x)\,J_{+\mu}(0)$, where
$J_\pm^\mu=\sum_q e_q\,\bar q_\pm \gamma^\mu q_\pm$.

\end{appendix}

\end{document}